\pdfoutput=1
\documentclass[cernpreprint, atlasdraft=false, UKenglish, texmf, orcidlogo]{atlasdoc}
\usepackage{atlaspackage}
\usepackage{atlasbiblatex}

\usepackage{atlasphysics}
\graphicspath{{logos/}{figures/}}

\usepackage{ANA-HMBS-2024-50-PAPER-defs}
\usepackage{makecell}

\AtlasTitle{Search for dark photons from Higgs boson decays in the gluon--gluon fusion channel in proton--proton collisions at $\sqrt{s}=13.6$ TeV with the ATLAS detector}

\AtlasVersion{2.0}

\AtlasAbstract{%
This paper presents a search for semi-visible Higgs boson decays in final states with a photon ($\gamma$) and missing transverse momentum $p_\mathrm{T}^{\mathrm{miss}}$.
The analysis is optimised for the gluon–-gluon fusion production channel and exploits a novel trigger deployed in 2023, with reduced thresholds down to 50 GeV on the photon transverse momentum ($p_\mathrm{T}^{\gamma}$) and 70 GeV on the $p_\mathrm{T}^{\mathrm{miss}}$,
in combination with a selection on the transverse mass of the $p_\mathrm{T}^{\mathrm{miss}}$–$\gamma$
system ($m_{\mathrm{T}}$) above 80 GeV. The results are interpreted in terms of Higgs boson decays into a $\gamma$ and a dark photon $\gamma_\mathrm{d}$ ($H\to\gamma\gamma_d$).
The analysis is performed using $135~\mathrm{fb}^{-1}$ of Run~3 data collected by ATLAS in 2023 and 2024. A boosted decision tree is used to identify events with misreconstructed primary vertices, which can affect the $p_\mathrm{T}^{\mathrm{miss}}$ calculation. Data-driven methods are employed to estimate the backgrounds from jets and electrons misidentified as photons, while the backgrounds with genuine photons are taken from Monte Carlo simulations, normalised to data in control regions with muons.
A binned maximum-likelihood fit to the $m_{\mathrm{T}}$ distribution is performed to evaluate the compatibility of the observed data with the Standard Model expectation. No significant excess is observed over the Standard Model prediction, and an observed (expected) upper limit on the branching ratio of $H\to\gamma\gamma_d$ of $1.4\%$ ($1.2\%$) is obtained at 95\%~confidence level. Combining this result with previous Run~2 searches improves the observed (expected) upper limit to $0.9$\% ($0.9$\%).
}

\AtlasRefCode{HMBS-2024-50}

\PreprintIdNumber{CERN-EP-2026-139}

\AtlasJournal{JHEP}
\AtlasJournalRef{JHEP 10 (2026) 270}
\AtlasDOI{10.1007/JHEP10(2026)270}

\hypersetup{pdftitle={ATLAS document},pdfauthor={The ATLAS Collaboration}}

\begin{document}

\maketitle

\tableofcontents

\section{Introduction}
%

% The next lines are included from the .//sections/intro.tex input file
%
\newcommand{\AtlasCoordFootnote}{%
ATLAS uses a right-handed coordinate system with its origin at the nominal interaction point (IP)
in the centre of the detector and the \(z\)-axis along the beam pipe.
The \(x\)-axis points from the IP to the centre of the LHC ring,
and the \(y\)-axis points upwards.
Polar coordinates \((r,\phi)\) are used in the transverse plane,
\(\phi\) being the azimuthal angle around the \(z\)-axis.
The pseudorapidity is defined in terms of the polar angle \(\theta\) as \(\eta = -\ln \tan(\theta/2)\) and is equal to the rapidity
$ y = \frac{1}{2} \ln \left( \frac{E + p_z}{E - p_z} \right) $ in the relativistic limit.
Angular distance is measured in units of \(\Delta R \equiv \sqrt{(\Delta y)^{2} + (\Delta\phi)^{2}}\).}

Exotic decays of the Higgs boson are well motivated by a variety of models beyond the Standard Model (BSM)~\cite{Curtin_2014}, including supersymmetry, dark sector or extended Higgs sectors, and provide a unique opportunity to probe new physics.

This paper presents a new strategy to search for exotic Higgs boson decays into a final state with a photon ($\gamma$) and missing transverse momentum (\ETmiss), probing a region of phase space with low kinematic thresholds that was previously unexplored by ATLAS.
This semi-visible decay of the Higgs boson provides an interesting production channel for the dark photon ($\gammad$), which arises as the gauge boson of an additional \( U(1)_D \) gauge group in dark sector models. Strong motivations for searches for this particle arise from its potential role as a ``portal'' to the dark sector, as well as from its possible contribution to addressing small-scale structure formation problems~\cite{Fan:2013tia,Heikinheimo:2015kra} and enabling viable asymmetric DM scenarios~\cite{Zurek:2013wia}.
In addition, this signature can arise in several BSM models, such as those with very low-scale supersymmetry breaking where the Higgs boson can decay into a gravitino ($\tilde{G}$) and a bino ($\tilde{B}$) that subsequently decays via $\tilde{B}\to\gamma\tilde{G}$~\cite{Petersson_2012}, in the Peccei--Quinn limit of next-to-minimal-supersymmetric models (NMSSM)~\cite{Suematsu_1998}, or in the decay of the Higgs boson into a pair of right-handed neutrinos ($N$) subsequently decaying via $N\to\nu\bar{\nu}\nu$ and $N\to\gamma \nu$~\cite{graesser2007}.

\begin{figure}[htbp]
\centering
\subfloat[]{\includegraphics[width=0.4\textwidth]{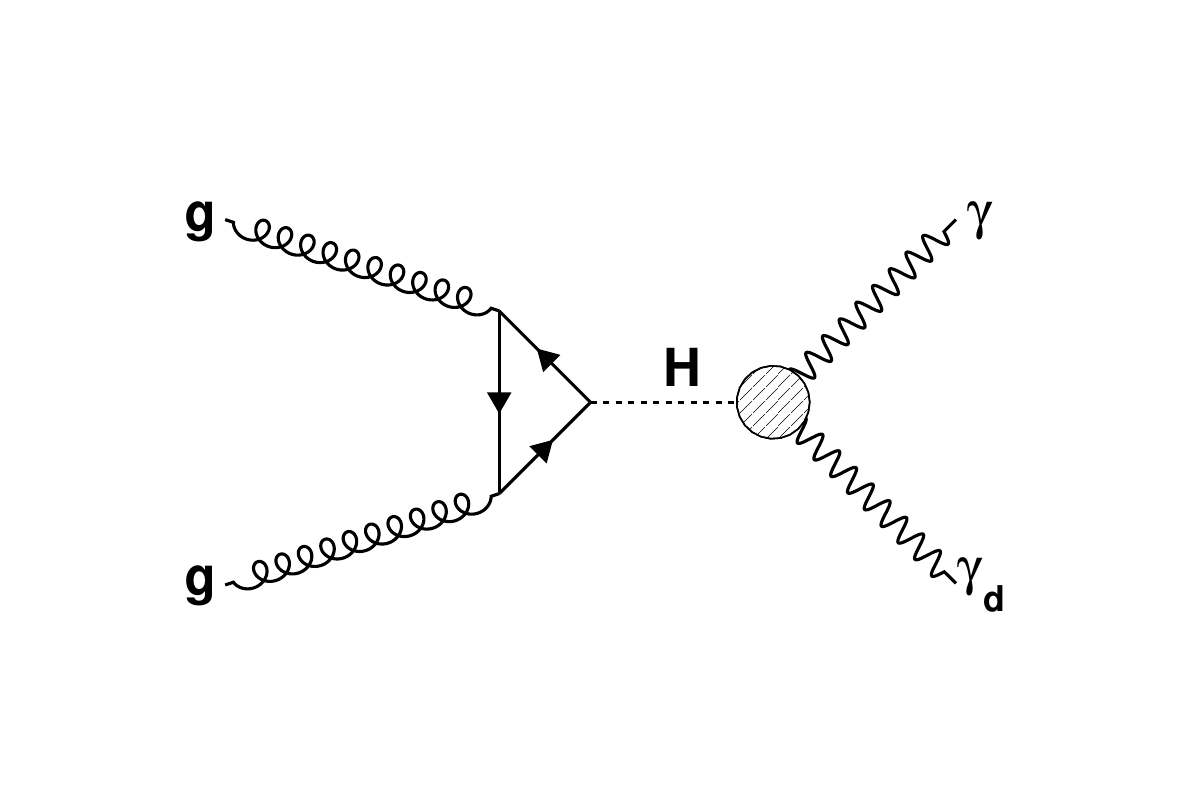}\label{ggh-diagram}}

\subfloat[]{\includegraphics[width=0.4\textwidth]{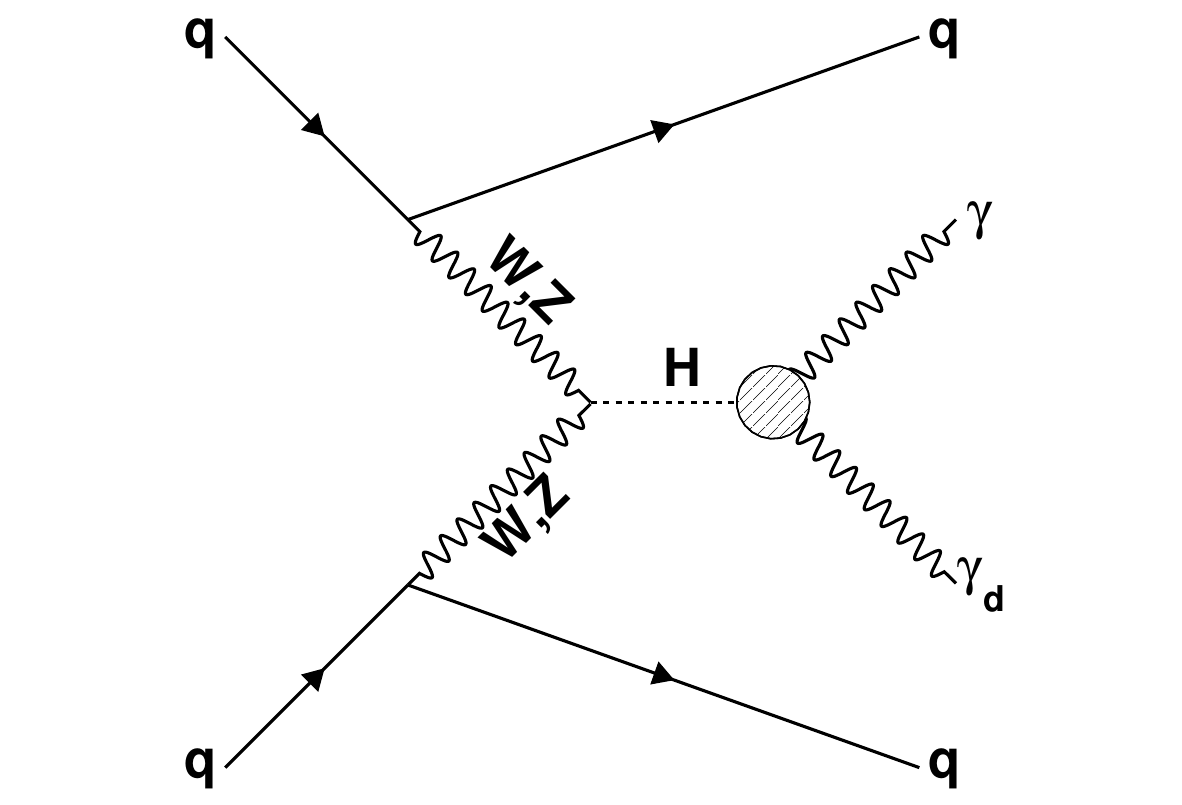}\label{vbf-diagram}}
\hspace{.1\textwidth}
\subfloat[]{\includegraphics[width=0.4\textwidth]{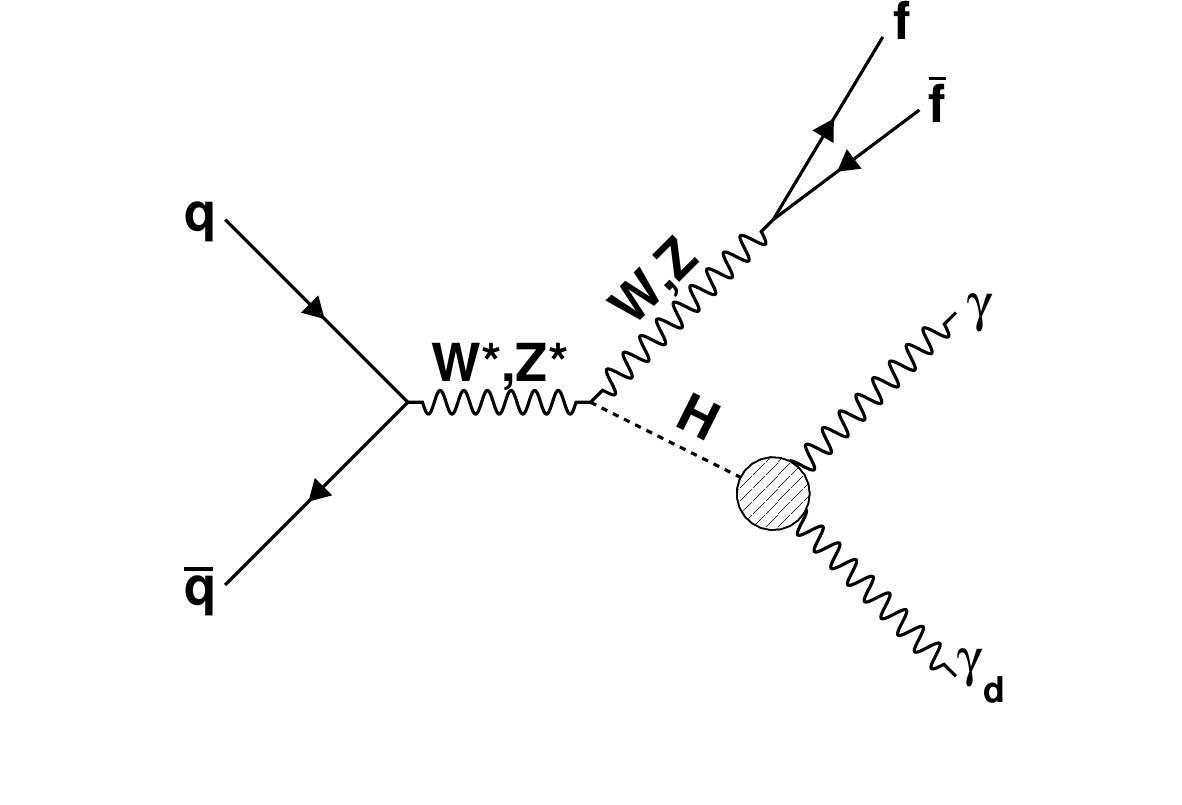}\label{vh-diagram}}
\caption{Feynman diagrams illustrating Higgs boson production via (a) the gluon--gluon fusion, (b) vector-boson fusion and (c) Higgsstrahlung processes, all followed by the decay into a photon and a dark photon ($H\to\gamma\gammad$). The hatched circle in the $H\gamma\gammad$ vertex represents the interaction through messenger fields.}
\label{fig:Feynman-all-Hyyd}
\end{figure}

While \gammad\ searches typically target the massive \gammad~\cite{Beacham_2019,harris2022snowmasswhitepapernew,Deliyergiyev_2016}, which can couple directly to the Standard Model (SM) via kinetic mixing with the ordinary photon~\cite{Fabbrichesi_2021}, this paper focuses on a massless \gammad\ scenario. In this case, the kinetic mixing between the \gammad\ and the photon can be tuned away through a rotation of the gauge fields, and the \gammad\ can interact with ordinary matter only through operators of dimension higher
than four~\cite{Fabbrichesi_2021}. The Higgs boson can decay through a loop interaction involving additional scalar or pseudo-scalar BSM messenger fields, which couple both to the SM and the dark sector~\cite{Gabrielli:2014oya, Biswas:2016jsh}, represented as an effective vertex in Figure~\ref{fig:Feynman-all-Hyyd}. Due to the non-decoupling properties of the Higgs boson, a branching ratio of $H\to\gamma\gammad$ (\br) up to $\sim 0.4\%$ may be allowed~\cite{Beauchesne2023a, Beauchesne2023b}, depending on the parameters of the model and taking into account constraints from searches for invisible Higgs boson decays and measurements of the branching ratio of $H\to\gamma\gamma$.

Searches for the semi-visible decay $H\to\gamma\gammad$  have been conducted at the LHC by the ATLAS and CMS collaborations, in the vector boson fusion (VBF) Higgs boson production channel and in associated production with a $Z$ boson ($ZH$)~\cite{HDBS-2019-13, EXOT-2021-17, CMS-EXO-19-007,CMS-EXO-20-005}. The combination of ATLAS results in the $ZH$ and VBF channels ~\cite{EXOT-2023-01} provides the best limit so far, excluding a $\br> 1.3\%$ (1.5\% expected) at 95\% confidence level (CL) with 139~fb\(^{-1}\) of data collected during the LHC Run~2 at $\sqrt{s}=13$\ TeV. The CMS results provide independent constraints, with an observed (expected) 95\% CL upper limit of $\br>2.9\%$ (2.1\%), obtained from the combination of the VBF and the $ZH$ channel~\cite{CMS-EXO-20-005}.
These production modes take advantage of a richer final state in addition to the $\gamma+\ETmiss$ signature from the $H\to\gamma\gammad$ decay, with additional jets or leptons that provide powerful handles for triggering, background rejection and identification of the hard-scatter primary vertex. In contrast, the gluon--gluon fusion mode (ggF), leading to the $\gamma+\ETmiss$ final state, is the dominant production mode, with a cross-section about an order of magnitude larger than that of the VBF channel. This final state has been explored by CMS ~\cite{CMS-HIG-14-025} and ATLAS in the context of other exotic models~\cite{EXOT-2018-63} or by focusing on the decay into $\gamma\gammad$ of BSM Higgs bosons with masses higher than 400~\GeV~\cite{ATL-PHYS-PUB-2023-003}. However, the $H \to \gamma \gammad$ decay of a Higgs boson with a mass of 125~\GeV has not been specifically targeted, as the low thresholds on the photon transverse momentum (\pTy) and \ETmiss\  required for this signature constitute a challenge for the trigger. Developments in the ATLAS trigger system between Run~2 and Run~3 led to the implementation of a composite trigger that combines selections on \ETmiss, \pTy, and the transverse mass between the $\gamma$ and the \ETmiss, $\mT=\sqrt{2\pTy\ETmiss(1-\cos(\Delta\phi))}$,\footnote{\AtlasCoordFootnote} where $\Delta\phi$ is the azimuthal separation between the photon and the \ETmiss. This new trigger allows to increase the acceptance of ggF events by approximately a factor of two, while keeping the trigger rate under control. By exploiting this new trigger, the analysis presented in this paper is optimised for the $H\to\gamma\gammad$ decay in the ggF production channel for the first time in ATLAS, while also retaining sensitivity to the VBF, $ZH$ and $WH$ channels.

% End of text imported from the .//sections/intro.tex input file

\section{ATLAS detector}
%

% The next lines are included from the .//sections/detector.tex input file
%

%
The ATLAS detector~\cite{PERF-2007-01,GENR-2019-02} at the LHC consists of an inner tracking detector system (ID) surrounded by a thin superconducting solenoid, electromagnetic and hadronic calorimeters (ECAL and HCAL),
and a muon spectrometer (MS) incorporating three large superconducting air-core toroidal magnets, with
an almost complete solid angle coverage around the collision point.

The ID system is immersed in a \qty{2}{\tesla} axial magnetic field
and provides charged-particle tracking in the range of \(|\eta| < 2.5\).
The high-granularity silicon pixel detector covers the interaction region and typically provides four measurements per track,
the first hit generally being in the insertable B-layer installed before Run~2~\cite{ATLAS-TDR-19,PIX-2018-001}.
It is followed by the silicon microstrip tracker, which usually provides eight measurements per track.
The outermost part of the ID is the transition radiation tracker (TRT),
which enables radially extended track reconstruction up to \(|\eta| = 2.0\).
The TRT also provides electron identification information
based on the fraction of hits (typically 30 in total) above a higher energy-deposit threshold corresponding to transition radiation.

The calorimeter system covers the pseudorapidity range \(|\eta| < 4.9\).
The ECAL~\cite{ATLAS-TDR-02,LARG-2009-01} is a lead/liquid-argon (LAr) sampling calorimeter covering the region \(|\eta|< 3.2\), divided into a barrel and two endcaps, and longitudinally segmented into three layers: the strips, the middle and the back layer. The 3-dimensional high-granularity in the fiducial acceptances $|\eta|<1.37$ and $1.52<|\eta|<2.37$ allows observation of the shape of the electromagnetic showers. Before the ECAL, a thin LAr presampler covering \(|\eta| < 1.8\) detects energy losses of showers starting in the material upstream.
The HCAL consists of a steel/scintillator-tile calorimeter,
segmented into three barrel structures within \(|\eta| < 1.7\), and two copper/LAr hadron endcap calorimeters.
The solid angle coverage is completed with forward copper/LAr and tungsten/LAr calorimeter modules
optimised for electromagnetic and hadronic energy measurements respectively.

The MS comprises separate trigger and
high-precision tracking chambers measuring the deflection of muons in a magnetic field generated by the superconducting air-core toroidal magnets.
The field integral of the toroids ranges between \num{2.0} and \qty{6.0}{\tesla\metre}
across most of the detector.
Three layers of precision chambers, each consisting of layers of monitored drift tubes, cover the region \(|\eta| < 2.7\),
except in the innermost layer of the endcap region, where layers of small-strip thin-gap chambers and Micromegas chambers provide precision tracking in the region \(1.3 < |\eta| < 2.7\).
The muon trigger system covers the range \(|\eta| < 2.4\) with resistive-plate chambers in the barrel, thin-gap chambers in the endcap regions,
and the small-strip thin-gap chambers and Micromegas chambers in the innermost layer of the endcap.

Interesting events are selected by the first-level (L1) trigger system implemented in custom hardware,
followed by selections made by software-implemented algorithms in the high-level trigger (HLT)~\cite{TRIG-2022-01}.
The first-level trigger selects events at a rate below \qty{100}{\kHz} from the \qty{40}{\MHz} bunch crossings; the high-level trigger further reduces that rate, in order to record events to disk at about \qty{3}{\kHz}.
An extensive software suite~\cite{SOFT-2022-02} is used in data simulation, in the reconstruction and analysis of real
and simulated data, in detector operations, and in the trigger and data acquisition systems of the experiment.

% End of text imported from the .//sections/detector.tex input file

\section{Data and simulation}
%

% The next lines are included from the .//sections/samples.tex input file
%

This analysis uses proton--proton (\(pp\)) collision data at a centre-of-mass energy of 13.6~\TeV, collected in 2023 and 2024 by the ATLAS detector. Only data recorded during stable beam conditions and with all sub-detectors operational are considered ~\cite{DAPR-2018-01}, corresponding to a total integrated luminosity of $135~\mathrm{fb}^{-1}$, with an uncertainty of 1.9\%~\cite{ATL-DAPR-PUB-2025-001}. Events were selected by using a dedicated trigger (`analysis trigger') deployed in 2023, optimised for the $\gamma+\met$ final state and described in Section~\ref{sec:trigger}.
In addition, samples with \ee, \ye\ and \ymu\ final states, used for the modelling of the background from electrons misidentified as photons (Section~\ref{sec:fey}) and for the efficiency studies of the analysis trigger (Section~\ref{sec:trigger}),
were selected using dielectron, single-electron, and single-muon triggers, respectively.

Monte Carlo (MC) simulations are used to model the signal and background processes described in Section~\ref{sec:signature}. In particular, the backgrounds containing genuine photons ($Z\gamma,~W\gamma$, and processes with a direct photon, \yjetdirect) are estimated using MC simulations.
The remaining backgrounds are estimated with data-driven methods, with $W$+jets and $Z$+jets MC samples used as input for the data-driven estimation of the background arising from jets misidentified as photons (Section~\ref{sec:fjy}).
All the samples, including processes with photons from fragmentation, \yjetfrag, and multijets samples, are used in the training of the algorithm that identifies wrongly reconstructed hard scattering vertices (Section~\ref{sec:sel-baseline}).
Simulations of events from top-quark production are not used, as the analysis applies a veto on jets emanating  from $b$-quarks (Section~\ref{sec:sel-baseline}).

The simulation of the geometry and response of the ATLAS detector~\cite{SOFT-2010-01} was performed using \GEANT~\cite{Agostinelli:2002hh}, and all simulated events were processed through the same reconstruction chain as collision data. The effect of multiple interactions in the same and neighbouring bunch crossings, pile-up (PU),
was modelled by overlaying~\cite{SIMU-2020-01} the simulated hard-scattering event
with inelastic \(pp\) events generated using \EPOS[2.0.1.4]~\cite{Werner:2005jf} and \PYTHIA[8.308]~\cite{Bierlich:2022pfr}.
The \EPOS events were generated with the \EPOS LHC set of tuned parameters (tune)~\cite{Pierog:2013ria}
and the \PYTHIA events with the A3 tune~\cite{ATL-PHYS-PUB-2016-017}
and the \NNPDF[2.3lo]~\cite{Ball:2012cx} set of parton distribution functions (PDF).
\PYTHIA PU events include either a high \pT jet,
a prompt photon, or a lepton from a $b$-hadron decay,
while events generated using \EPOS were filtered to simulate all remaining PU events in the overlay sample.
The individual simulations were first reweighted to ensure a smooth jet \pt\ distribution. The combination was then reweighted to match the distribution of the actual number of interactions per bunch crossing measured in data.

The signal samples include multiple Higgs boson production channels. The dominant process and main target of this analysis is the ggF channel, but non-negligible contributions arise from the VBF, $WH$ and $ZH$ channels. One MC sample for each production mode was generated, with $WH$ production divided into $W^+H$ and $W^-H$. The matrix elements were estimated
using \POWHEG[v2]~\cite{Alioli:2010xd,Frixione:2007vw} at next-to-leading order (NLO), with the PDF4LHC21 set of PDFs~\cite{Ball:2022oua}.
For the ggF channel, next-to-next-to-leading order (NNLO) precision was achieved through a reweighting of the events computed in bins of the \pT\ and rapidity of the generated Higgs boson, to match NNLOPS predictions~\cite{Hamilton:2013fea}. The samples were generated with the SM Higgs boson mass set to 125~\GeV.
All the signal samples were normalised to state-of-the-art cross-section predictions for the centre-of-mass energy of 13~\TeV~\cite{deFlorian:2016spz}, extrapolated to 13.6~\TeV~\cite{karlberg2024adinterimrecommendationshiggs}.
The decay into photons and dark photons, as well as the hadronisation and showering, was simulated using \Pythia[8.245] with the CTEQ6L1 PDF set and the AZNLO tune~\cite{Pumplin:2002vw}. The `Hidden Valley' scenario~\cite{Carloni_2010} was used for modelling the $H\to\gamma\gammad$ decay.

The $V\gamma$ and $V+$jets processes were simulated using the \Sherpa[2.2.14] generator~\cite{Bothmann:2019yzt}, with up to one jet (two jets) at NLO and up to three additional jets at leading order (LO) for $Z\gamma$ and $W\gamma$ ($Z+$jets and $W+$jets). The matrix elements were matched and merged with the \Sherpa parton shower~\cite{Schumann:2007mg}, using the MEPS@NLO prescription~\cite{Hoeche:2011fd,Catani:2001cc} and the NNPDF3.0 NNLO PDF set~\cite{Ball:2014uwa} with the dedicated \Sherpa\ tune.
The \yjetdirect, \yjetfrag\ and multijets samples were generated at LO using \Pythia[8.186]~\cite{Mrenna:2016sih,Sjostrand:2007gs} with the NNPDF2.3 LO PDF set and the A14 tune~\cite{ATL-PHYS-PUB-2014-021} for the parton shower.
Table~\ref{tab:mc_summary} summarises the configurations used for the signal and background MC simulations of the analysis.

\begin{table}[]
\centering
\footnotesize
\caption{Configurations used for event generation and parton shower of signal (ggF, VBF, $ZH$ and $WH$) and background MC samples. The matrix element (ME) order, leading-order (LO) or next-to-leading order (NLO), refers to the order in the strong coupling constant of the perturbative calculation in the MC event generation. PDF refers to the parton density libraries used with the generator. The tune refers to the underlying-event tune of the parton shower model.}
\label{tab:mc_summary}
\begin{tabular}{lccccc}
\toprule
Process & ~~Generator~~ & ~~ME order~~ & ~~PDF~~ & ~~ Parton Shower ~~& ~~Tune~~ \\
\midrule
$H\to\gamma\gammad$ & \POWHEG[v2] & NLO & PDF4LHC21 & \Pythia[8.245] & AZNLO \\
$Z\gamma, W\gamma$ & \Sherpa[v2.2.14] & 0,1j@NLO + 2,3,4j@LO & NNPDF3.0 NNLO & \Sherpa & \Sherpa  \\
$Z$+jets, $W$+jets & \Sherpa[v2.2.14] & 0,1,2j@NLO + 3,4,5j@LO & NNPDF3.0 NNLO & \Sherpa & \Sherpa \\
\yjetdirect & \Pythia[8.186] & LO & NNPDF2.3 LO & \Pythia[8.308] & A14 \\
\yjetfrag & \Pythia[8.186] & LO & NNPDF2.3 LO & \Pythia[8.308] & A14 \\
multijets & \Pythia[8.186] & LO & NNPDF2.3 LO & \Pythia[8.308] & A14 \\
\bottomrule
\end{tabular}
\end{table}

% End of text imported from the .//sections/samples.tex input file

\section{Analysis trigger}
\label{sec:trigger}

% The next lines are included from the .//sections/trigger.tex input file
%

The `analysis trigger' combines selections on the photon \pT, the \ETmiss\ and the \mT. It selects 10\% of the total generated signal events in the $ggF$ channel, with a factor of two increase relative to previous single-photon or inclusive \pTmiss\ triggers. A similar trigger efficiency is observed also in the VBF and $VH$ channels. At L1, the trigger selected electrons or photons with \pT\ above 26~\GeV. The HLT further required a %
50~\GeV photon fulfilling tight identification criteria~\cite{TRIG-2022-01}. %
To compute the \pTmiss\ at HLT level, a two-stage sequence of algorithms was employed, starting with the `cell' \pTmiss\ computed using only the calorimeter cells as input~\cite{TRIG-2019-01}. This fast step was followed by a more sophisticated \pTmiss\ algorithm using particle flow~\cite{PERF-2015-09} constituents, combining tracks and calorimeter clusters, with a dedicated PU subtraction algorithm (`pfopufit')~\cite{TRIG-2022-01}. A minimum threshold of 40~\GeV was applied for the first `cell' \pTmiss\ requirement, while the second `pfopufit' step had a threshold of 70~\GeV.
To further reduce the rate, the trigger required a minimum \mT\ (computed using `pfopufit' \pTmiss) of 80~\GeV.
Since muons are invisible to the `cell' \pTmiss\ algorithm~\cite{TCAL-2010-01}, particle flow constituents consistent with muons were removed also in the `pfopufit' algorithm, to preserve the same behaviour. Therefore both the `cell' and the `pfopufit' algorithms were insensitive to the presence of muons, which were treated as invisible in the computation of \pTmiss. This feature allowed samples to be selected with reconstructed muons in the final state and to treat them as invisible to mimic \pTmiss, without introducing a bias induced by the analysis trigger. In particular, events with muons are used to constrain the $W\gamma$ and $Z\gamma$ backgrounds in control regions with one or two muons as described in Section~\ref{sec:bkg}.

\begin{figure}[ht]
\centering
\subfloat[]{\includegraphics[width=0.40\linewidth]{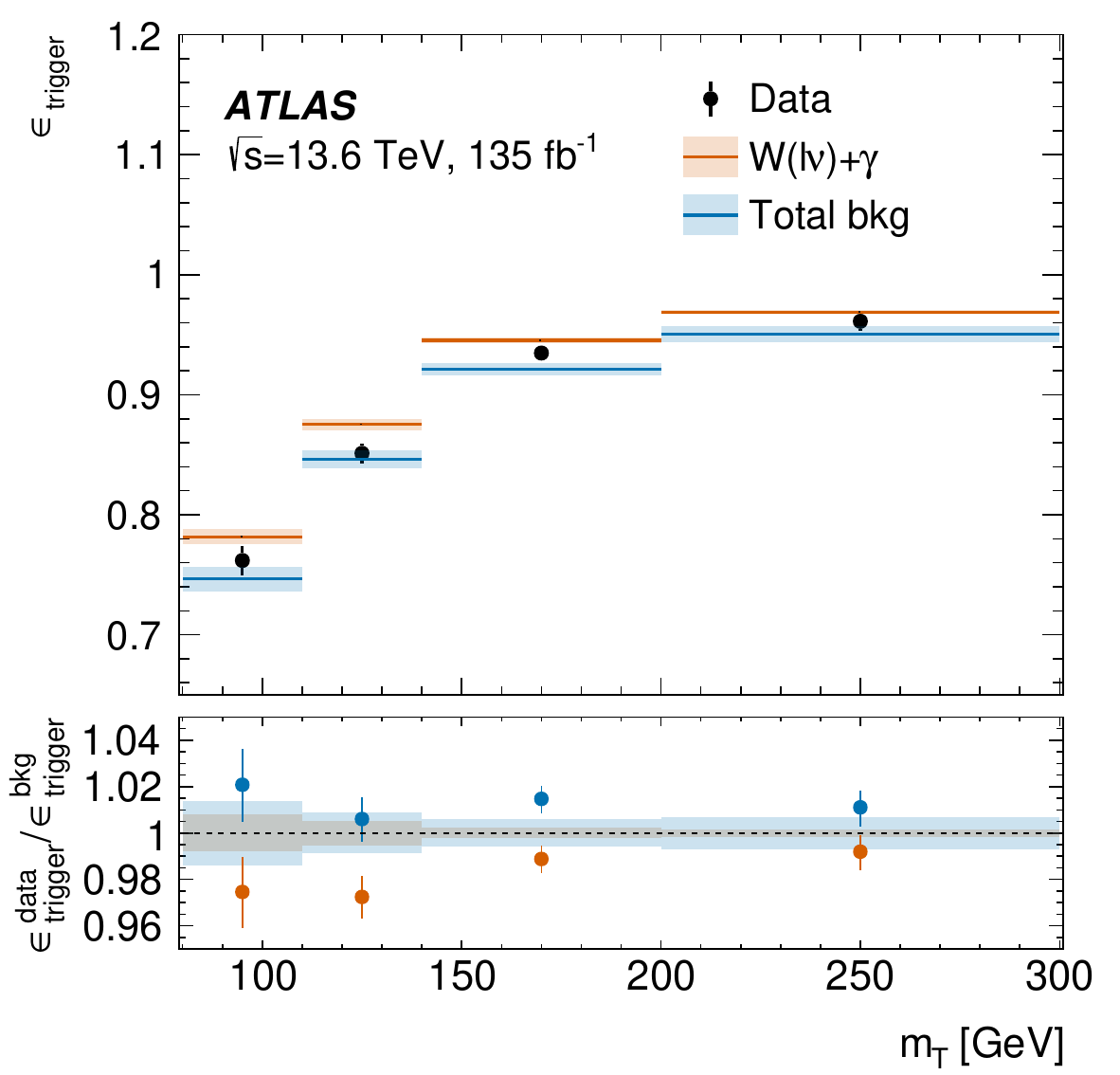}\label{fig:trig_mt}}\\
\subfloat[]{\includegraphics[width=0.40\linewidth]{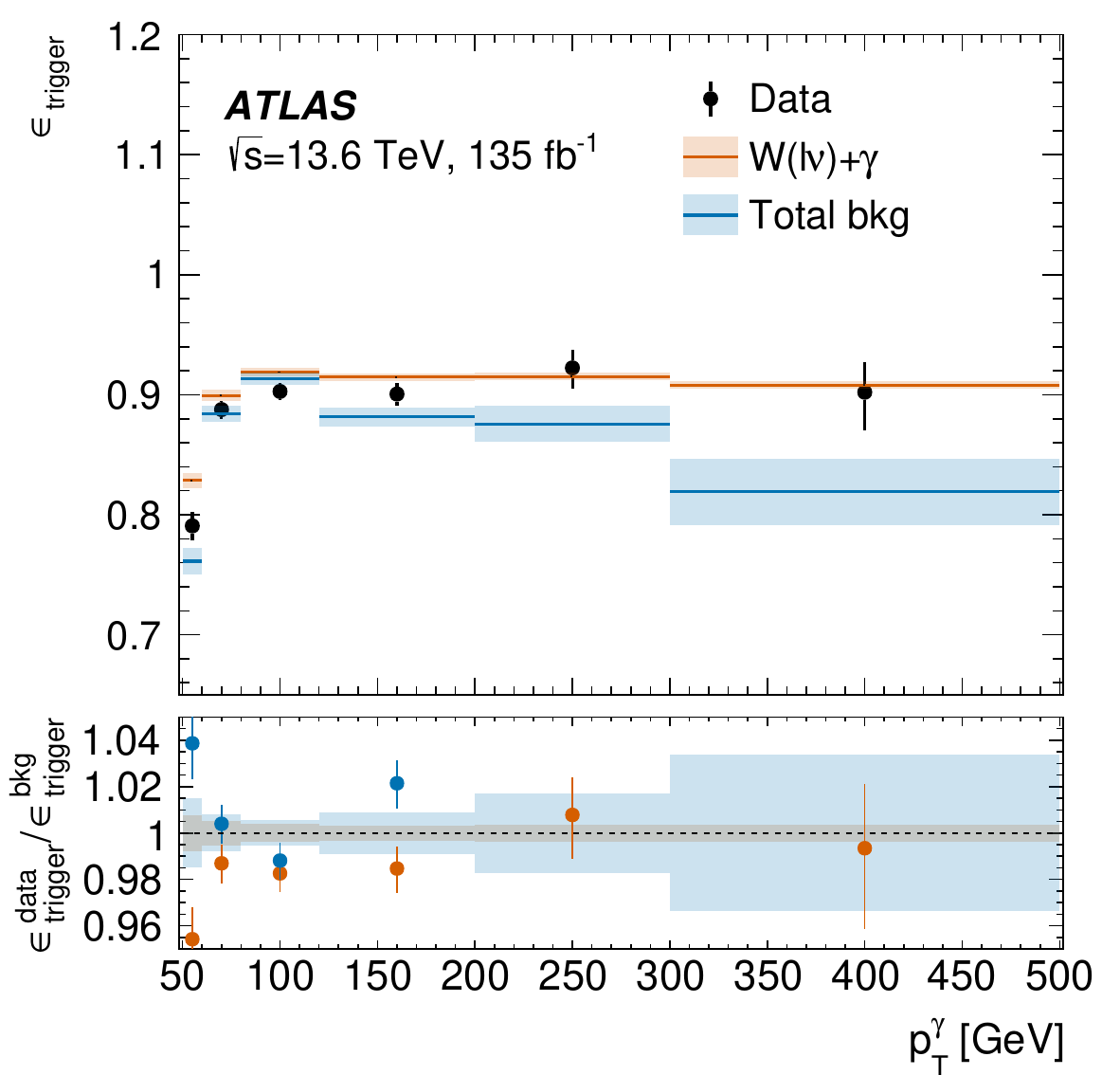}\label{fig:trig_phpt}}
\subfloat[]{\includegraphics[width=0.40\linewidth]{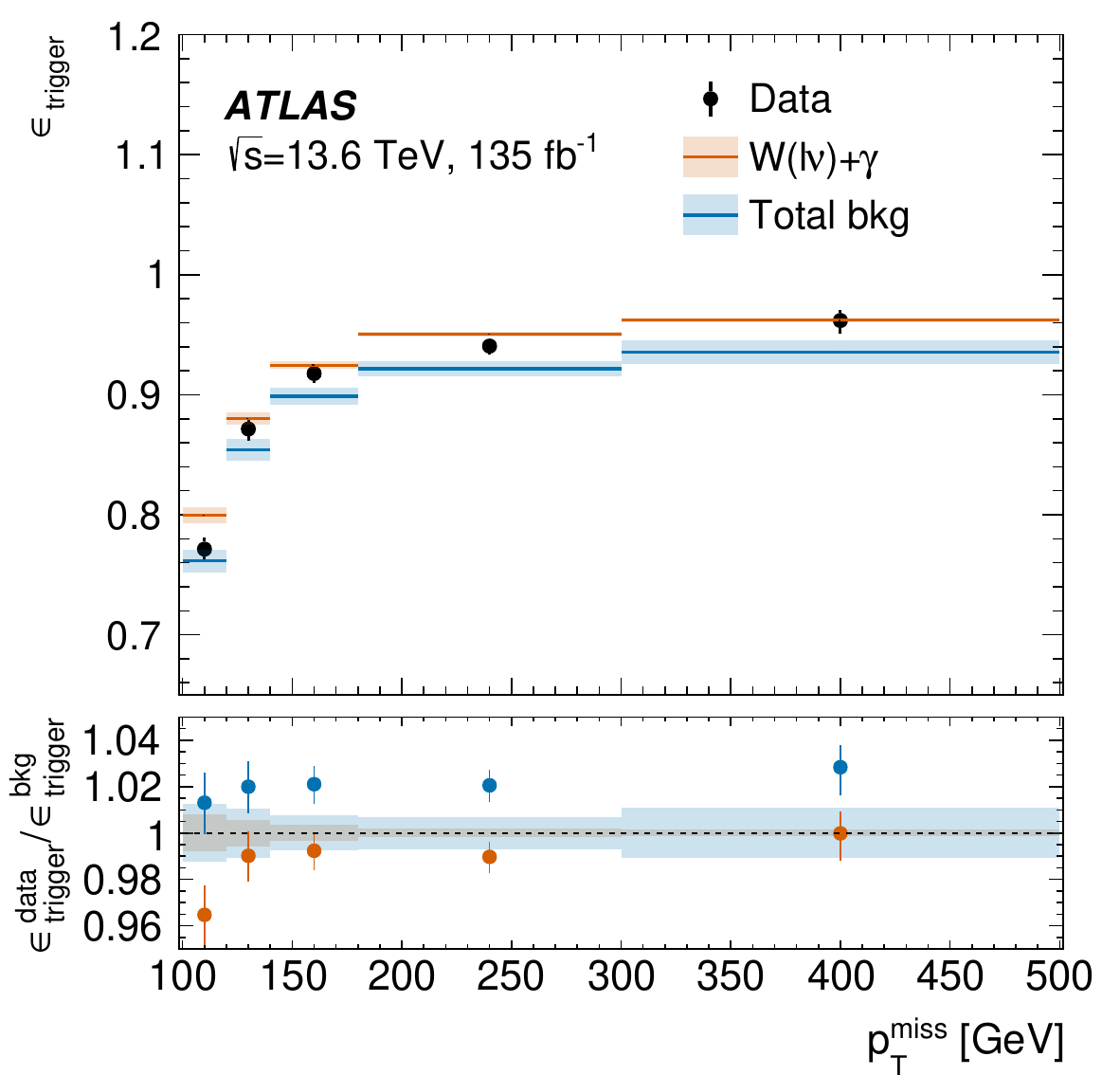}\label{fig:trig_met}}
\caption{Analysis trigger efficiency as a function of (a) \mT\, (b) \pTy\ and (c) \pTmiss\, for data, simulation of the dominant background $W\gamma$ and total background. Events were selected using single-muon triggers and requiring one muon treated as invisible in the evaluation of \pTmiss\ and \mT. In addition, a subset of the selections used in the analysis is applied: $\ETmiss>100$~\GeV, $\mT>80$~\GeV, one photon with $\pTy>50$~\GeV and a requirement of $\dphimety>1.25$ on the azimuthal separation between the \pTmissvec\ and its photon term $\vec{p}_\mathrm{T}^{\mathrm{miss},\gamma}$, equivalent to $-\pTyvec$. The bottom panels show the ratios of the efficiency in data to those in the $W\gamma$ MC simulation and in the total background expectation. The shaded bands and the error bars represent the statistical uncertainties associated with the MC simulation and the data, respectively.}
\label{fig:trigger_eff}
\end{figure}

Events with one muon are also used to assess the accuracy of the simulation of the analysis trigger in \ymu\ final states, selected by a muon trigger used as a baseline trigger. The muon was treated as invisible in the computation of \pTmiss\ and \mT\ in the full (offline) reconstruction.
Figure~\ref{fig:trigger_eff} shows the trigger efficiency as a function of offline \pTy, \pTmiss\ and \mT. The efficiency in data is compared to that of the dominant background contribution from $W\gamma$ events (75\%), and that of the total background, which includes contributions from $Z\gamma$ events (2\%) and from events with a jet misidentified as a photon (23\%). The latter may exhibit a different trigger efficiency due to an expected different trigger response to misidentified jets.
The trigger efficiency measured in data is observed to lie in between the one predicted by the simulations for $W\gamma$ background and for the total background. Up and down uncertainties in the trigger efficiency are therefore assigned as the differences between the efficiency measured in data and the one predicted for the total background and for the $W\gamma$ simulation, respectively. This procedure also accounts for the uncertainty in the modelling of the background from jets misidentified as photons. The \mT\ variable is correlated with both the \pTy\ and the \pTmiss.  While the trigger efficiency curve rapidly reaches a plateau as a function of \pTy, a slower turn-on is observed as a function of \pTmiss. For this reason, uncertainties at the 2\%--3\% level are applied in bins of \pTmiss. Since the trigger efficiency for signal events is found to be consistent with that observed in the $W\gamma$ simulation, the same uncertainty is assigned to the signal sample.

% End of text imported from the .//sections/trigger.tex input file

\section{Event reconstruction}
\label{sec:object-rec}

% The next lines are included from the .//sections/object-rec.tex input file
%

For each event, vertex candidates are required to have at least two associated tracks with $\pT > 500$~\MeV. Among the reconstructed vertices, the hard-scattering (HS) primary vertex (PV) is selected as the one with the highest sum of squared transverse momenta of contributing tracks~\cite{IDTR-2021-01}.

For each reconstructed physics object, looser selection criteria are first applied to define \textit{baseline} objects, which are used for event vetos, \pTmiss\ reconstruction, and overlap removal. More stringent criteria are then required, to define \textit{selected} objects.

Electrons are reconstructed by matching energy clusters in the ECAL to tracks in the ID~\cite{EGAM-2021-02,EGAM-2021-01}. \textit{Baseline} electrons are required to satisfy $p_{\mathrm{T}}>10$~\GeV and $|\eta|<2.47$, be outside the transition region between the ECAL barrel and endcaps ($1.37< |\eta| < 1.52$), and be consistent with originating from the PV through requirements on the longitudinal and transverse impact parameters, $|z_0\sin(\theta)|<0.5$~mm and $|d_0/\sigma(d_0)|<5$ where $\sigma(d_0)$ is the uncertainty in $d_0$. All baseline electron candidates are required to satisfy \leploose\ likelihood-based identification criteria~\cite{PERF-2017-01}, based on calorimeter, tracking and combined variables that provide separation between electrons and jets. \textit{Selected} electrons are further required to satisfy \leptight\ identification and \leptight\ isolation requirements~\cite{PERF-2017-01}.

Muons are reconstructed by matching ID tracks to MS tracks or track segments, by associating ID tracks with calorimeter energy deposits compatible with a minimum-ionizing particle, or by identifying MS tracks
satisfying the \leploose\ requirement and compatible with the interaction point~\cite{MUON-2018-03,PERF-2015-10}. \textit{Baseline} muon candidates must satisfy $p_{\mathrm{T}}>10$~\GeV and $|\eta|<2.7$ and \lepmedium\ identification criteria. Similar selections on the impact parameter as for electrons are applied, but with $|d_0/\sigma(d_0)| < 3$. \textit{Selected} muons are further required to satisfy \leploose\ isolation selections~\cite{PERF-2015-10}.

Photon candidates are reconstructed from clustered energy deposits in the ECAL~\cite{EGAM-2021-02,EGAM-2021-01}.
For photons that undergo conversions in the material upstream the ECAL, tracks in the ID are also used.
\textit{Baseline} photons are required to
satisfy $\pTy>25$~\GeV, $|\eta|<2.37$ and be outside the transition region between the ECAL barrel and endcaps ($1.37< |\eta| < 1.52$). Photon candidates must satisfy a \T\ identification criterion, based on several requirements on the `shower shape' variables~\cite{PERF-2017-02}, which can be grouped into three categories: energy leakage in the HCAL, width in the middle layer of the ECAL, and shape in the strips.
\textit{Selected} photon candidates are additionally required to be isolated. A dimensionless isolation variable is introduced:
\begin{equation}
\isol=\dfrac{\ETiso-2.45~\GeV}{\pTy},
\label{eq:iso}
\end{equation}
where \ETiso\ is the transverse energy measured by the calorimeters in a cone of size $\Delta R=0.4$ around the photon candidate.
Isolated photons are required to satisfy the \TightCaloOnly\ working point, $\isol<0.022$, a typical requirement to select prompt, isolated photons in ATLAS~\cite{EGAM-2018-01}.
Non-isolated photons are used in the evaluation of the background induced by jets misidentified as photons (see Sections~\ref{sec:signature} and ~\ref{sec:fjy}).
In this context, a \Loose\ photon selection is required, with relaxed requirements on the HCAL and ECAL middle layer, and no requirements on strips. Removing \T\ photons from the \Loose\ sample provides a \Lp\ sample enriched in this background. Further photon selections $\Lpp$ are introduced as subsets of the $\Lp$ sample, by applying the same selections as for \T\ to the HCAL and middle layer shower shapes, but at most one requirement on one of the strips shower shapes.

Particle flow (PFlow) jets are reconstructed using the anti-$k_t$ algorithm~\cite{
Cacciari:2008gp}, as implemented in FastJet~\cite{
Fastjet}, with a radius parameter
$R = 0.4$. The algorithm takes PFlow objects as inputs, combining measurements from the ID and the calorimeters~\cite{PERF-2015-09}.
\textit{Selected} jets are required to satisfy $\pT>20~\GeV$.
In addition, central jets ($|\eta| < 2.5$) are classified as originating from the HS interaction using a neural-network-based jet-vertex-tagger (NN-JVT) (an evolution of the JVT described in Ref.~\cite{PERF-2014-03}), while a forward jet vertex tagging (fJVT) algorithm~\cite{PERF-2016-06} is applied to forward jets ($|\eta| > 2.5$).
Jets containing $b$-hadrons (\bjets) are identified using a graph neural network tagger with 85\% $b$-tagging efficiency~\cite{FTAG-2023-05}.
To suppress the background originating from detector noise, hardware issues, beam-induced background or cosmic-ray muons, events containing jets failing to satisfy \leploose\ jet-quality criteria are rejected~\cite{ATLAS-CONF-2015-029}.

Hadronically decaying $\tau$-leptons ($\tau_\mathrm{had}$) are seeded by anti-$k_t$ jets with $R=0.4$ and identified using a recurrent neural network algorithm that uses
tracks and calorimeter energy clusters associated with $\tau_\mathrm{had}$ candidates, as well as high-level discriminating
variables~\cite{ATL-PHYS-PUB-2019-033}. \textit{Baseline} $\tau_\mathrm{had}$ are required to satisfy a \lepmedium\ identification working point and have $\pT>20$~\GeV and $|\eta| < 2.5$.

The $\pTmissvec$, with magnitude $\pTmiss=|\pTmissvec|$, is defined as the negative vector sum of the transverse momenta of the \textit{baseline} reconstructed objects (`hard term'), and of the `soft term' estimated from charged PFlow objects associated with the primary vertex and not matched to any of the reconstructed hard objects~\cite{JETM-2020-03}. In the \pTmissvec\ calculation, a NN-JVT selection is applied to the jets for PU subtraction. An alternative \pTmissvec\ with no PU subtraction, referred to as \pTmissvecNoJVT, is also exploited to help tag events with a misidentified PV, as explained in Section~\ref{sec:selections}. In addition, a key quantity to identify events with fake \ETmiss\ is the \ETmiss\ significance
$\metsig=\frac{\ETmiss}{\sigma(\ETmiss)}$, where $\sigma(\ETmiss)$ denotes the $\ETmiss$ resolution, accounting for the resolution of the reconstructed physics objects entering the \ETmiss\ computation~\cite{ATLAS-CONF-2018-038,JETM-2020-03}.

After the reconstruction of all physics objects, a sequential overlap removal procedure, as detailed in Table~\ref{tab:or}, is applied to resolve ambiguities by keeping one single object in case of overlapping objects, based on $\Delta R$ matching, shared ID tracks or calorimeter energy clusters, according to a predefined priority.

\begin{table}[ht]
\caption{Overview of the sequential overlap removal procedure applied to reconstructed objects and the corresponding matching criteria based on the $\Delta R$ separation and shared ID tracks. If two objects overlap based on the matching criteria described in the third column, the object in the first column is removed while the one in the second column is retained. The table is ordered according to the object priority used in the overlap removal procedure.}
\label{tab:or}
\centering
\begin{tabular}{lll}
\toprule
Remove & Keep & Matching criteria \\
\midrule
Muon     	& Electron & Muon with calorimeter deposits and shared ID track \\
Electron 	& Muon     & Shared ID track \\
Jet      		& Electron & \(\Delta R < 0.2\) \\
Electron 	& Jet      & \(0.2\ < \Delta R < 0.4\) \\
Electron	& Electron & Shared ID track,  electron with lower \pT\ removed\\
Jet      		& Muon     & Number of tracks < 3 and \(\Delta R < 0.2\) \\
Muon 	    & Jet      & \(0.2\ < \Delta R < 0.4\) \\
$\tau$-lepton     	 	& Electron & \(\Delta R <\) 0.2 \\
$\tau$-lepton       	 	& Muon     & \(\Delta R <\) 0.2 \\
Jet      		& $\tau$-lepton      & \(\Delta R < 0.2\) \\
Photon   	& Electron & Shared calorimeter energy clusters \\
Photon   	& Muon     & \(\Delta R < 0.4\) \\
Jet   	& Photon      & \(\Delta R < 0.4\) \\
\bottomrule
\end{tabular}
\end{table}

% End of text imported from the .//sections/object-rec.tex input file

\section{Signature and backgrounds}
\label{sec:signature}

% The next lines are included from the .//sections/signature.tex input file
%

The $H\to\gamma\gammad$ signal is characterised by the presence of a single photon and \ETmiss. Although the $z$-component of the \gammad\, momentum cannot be reconstructed, the \mT\ distribution features a broad peak around the Higgs boson mass. In the ggF channel, the final state contains no leptons, while hadronic jets may originate from gluon radiation.
The same final state also covers the VBF channel and $VH$ production with hadronic decay of the vector bosons.
Due to the trigger thresholds on \ETmiss\ and photon \pTy\ (Section~\ref{sec:trigger}), optimised as a compromise between signal acceptance and trigger rate, the selected signal events are mostly Higgs bosons recoiling against jets, resulting in a boosted topology.

The analysis is significantly affected by background from jets and electrons misidentified as photons, as well as by events reconstructed with large \ETmiss, due to mismeasurement or misreconstruction of physics objects (referred to as \textit{fake \ETmiss}).
In addition, in this signature both the photon and the \pTmiss\ have no associated tracks, therefore the reconstruction of the PV is particularly challenging, since it relies exclusively on tracks originating from the underlying event and from additional jets. When the PV is misidentified, HS jets can be incorrectly classified as PU jets and consequently removed from \pTmiss\ computation, thus enhancing the contribution from fake \pTmiss\ background.
The SM background events with similar signatures can be classified in the following categories.
\begin{itemize}
\item \textbf{Genuine photon}: $Z(\to\nu\nu)\gamma$ (irreducible background) or $W(\to\ell\nu)\gamma$ and $Z(\to\ell\ell)\gamma$ where the lepton(s) $\ell$ are not reconstructed. An additional contribution arises from $\gamma+\mathrm{jets}$ processes with a direct photon, $\yjetdirect$, with fake \ETmiss\ originating primarily from lost or mismeasured jets.
\item \textbf{Electrons misidentified as photons (\efakey)}: mostly $W(\to e\nu)+\mathrm{jets}$ events.
\item \textbf{Jets misidentified as photons (\jfakey)}: dominantly multijets and $\gamma$+jets with a fragmentation photon ($\yjetfrag$) --- in the latter, the photon is considered as \jfakey, since it is embedded in a hadronic jet and therefore not isolated. Such events pass the analysis selections when associated with fake \ETmiss. A smaller contribution arises from $Z(\to\nu\nu)$+jets and $W(\to\ell\nu)$+jets processes.
\end{itemize}

% End of text imported from the .//sections/signature.tex input file

\section{Event selection}
\label{sec:selections}

% The next lines are included from the .//sections/selections.tex input file
%

Events satisfying the trigger selections described in Section~\ref{sec:trigger} are further required to satisfy a set of offline requirements, designed to isolate the signal.

\subsection{Baseline selection}
\label{sec:sel-baseline}

Events are required to have exactly one selected photon with $\pTy > 50$~\GeV and no additional \textit{baseline} photons, $\ETmiss > 100$~\GeV, a transverse mass $\mT > 80$~\GeV, no \textit{baseline} leptons nor \btagged\ jets, and at most three central jets. Jet cleaning requirements \cite{ATLAS-CONF-2015-029} are applied to reject events with bad quality jets, which may arise from electronic noise or non-collision background. The veto on the \btagged\ jets is applied to reduce the contribution from background processes with top quarks. %
The requirements on \pTy\ and \mT\ are aligned with the trigger thresholds, exploiting the fast trigger turn-on, while a more conservative requirement is applied to \ETmiss\  as a compromise between signal acceptance and background rejection.

\begin{figure}[ht]
\centering
\includegraphics[width=0.7\linewidth]{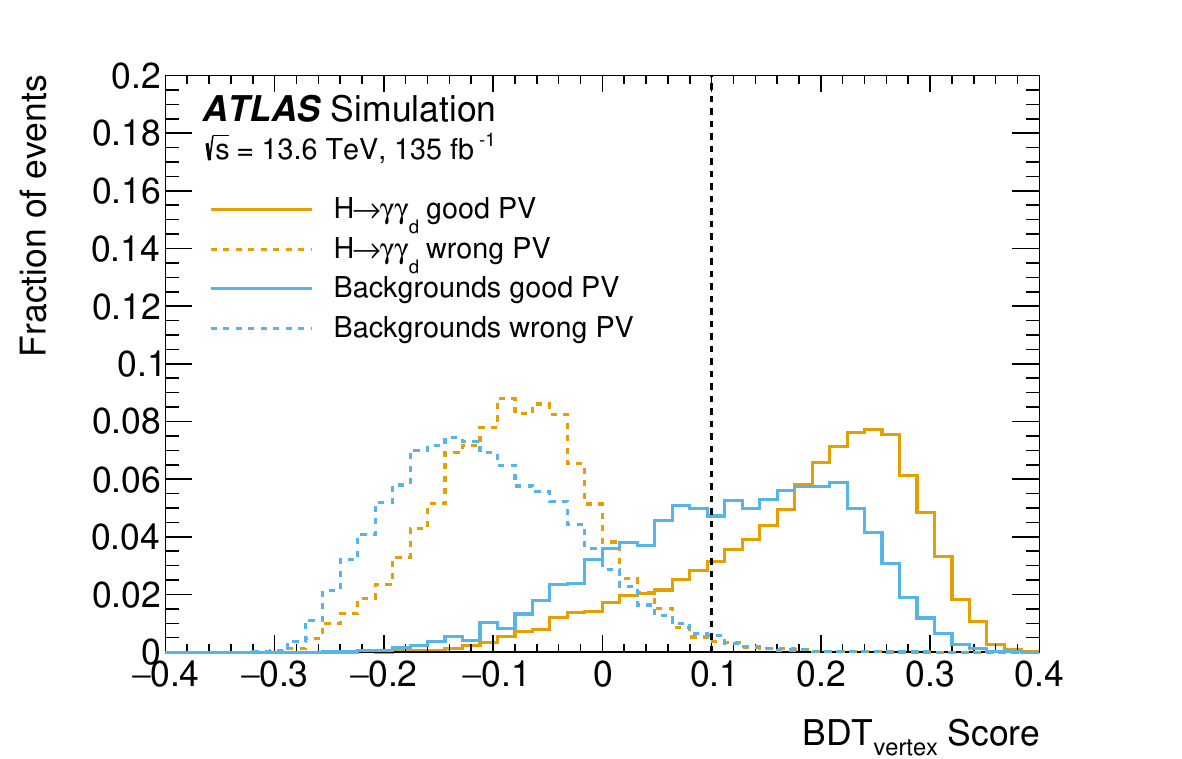}
\caption{Normalised distributions of the $\mathrm{BDT_{vertex}}$ score for events with correctly identified PV (solid line) and misidentified PV (dashed line). The distribution is shown for the sum of all signal channels and background processes. The vertical dashed line indicates the selection applied in the analysis.}\label{fig:bdt-distr}
\end{figure}

\begin{figure}[!htbp]
\centering
\includegraphics[width=0.7\linewidth]{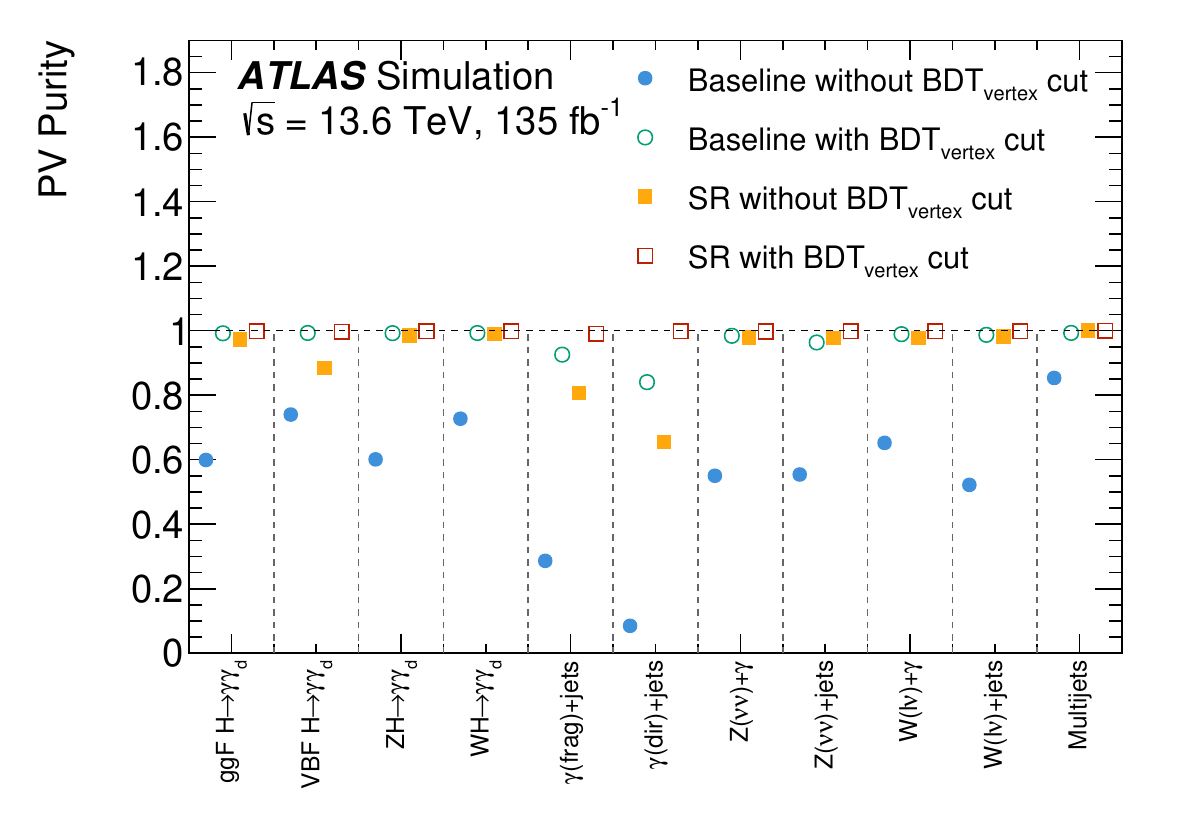}
\caption{ Fraction of events with the correct PV for the $H\to\gamma\gammad$ signals and for different background processes, with different selections applied: the baseline selection and the SR, with and without the $\mathrm{BDT}_{\mathrm{vertex}}$ requirement.}\label{fig:bdt}
\end{figure}

As discussed in Section~\ref{sec:signature}, low track activity can lead to an incorrect identification of the PV\@. This effect has a significant impact on \ETmiss-related observables, particularly when HS jets are misclassified as PU jets and consequently excluded from the \ETmiss\ reconstruction. In such events, the reconstructed \ETmiss\ distribution becomes artificially harder for $\gamma$+jets and multijets backgrounds and softer for the signal, enhancing background acceptance while reducing signal efficiency. Moreover, the PV misidentification distorts the transverse mass distribution, thereby degrading the signal-to-background discrimination, since the loss of HS jets results in a final state with one photon back-to-back with \pTmiss, for both signal and background events. This leads to values of \mT\ above 140~\GeV, reflecting the \pTy\ and \pTmiss\ thresholds of 50~\GeV and 100~\GeV, respectively.
To mitigate this effect, a boosted decision tree (BDT) is trained on signal and background simulations using TMVA AdaBoost~\cite{FREUND1997119}, to identify and reject events with a misidentified PV\@. This approach ensures a good control of the background modelling at the baseline selection level and, therefore, in all the analysis regions. The BDT uses information related to the reconstructed vertex properties, jet multiplicity, and the angular separation between the photon, jets, and \ETmiss, as summarised in Table~\ref{tab:bdt_features}. The vertex BDT also further suppresses the contribution from non-collision background events with a photon. In particular, the position of the vertex inferred from the photon trajectory in the calorimeter (photon pointing) is used as one of the input variables, favouring events in which the photon originates from the PV.
The distribution of the BDT score for events with correct and incorrect PV identification is shown in Figure~\ref{fig:bdt-distr} for signal and background processes separately, confirming a good discriminating power. As highlighted in Figure~\ref{fig:bdt}, a requirement on the misidentified vertex BDT score of $\mathrm{BDT}_{\mathrm{vertex}}>0.1$ provides an almost complete suppression of events with a misidentified PV for both signal and background, especially when combined with the signal region (SR) selection described below. The efficiency of this selection for signal events, relative to the other baseline requirements, is approximately 55\%, increasing to $\sim 95\%$ in the SR.
The $\mathrm{BDT}_{\mathrm{vertex}}$ distribution is validated in the main analysis regions defined below, as well as in a dedicated region with $\mT>200$~\GeV and no leptons, enriched with $W\gamma$ and $Z\gamma$ processes with a topology similar to that of the signal. In all cases, good agreement between data and the background expectation is observed for $\mathrm{BDT}_{\mathrm{vertex}}>0.1$, and the acceptance of this requirement is found to be consistent between data and the background prediction. Therefore, no systematic uncertainty is assigned to the $\mathrm{BDT}_{\mathrm{vertex}}$ selection.\\

\begin{table}[!htbp]
\centering
\renewcommand{\arraystretch}{1.5}
\small
\caption{Description of the input features of the misidentified vertex BDT, in order of importance.}
\label{tab:bdt_features}
\begin{tabular}{p{3.5cm} p{7.5cm}}
\toprule
\textbf{Variable} & \textbf{Description} \\
\midrule
$\log\!\left(\sum p_{\mathrm{T}}^{2}(\text{tracks})\right)$
&  Logarithm of the sum of the squared transverse momenta of tracks associated with the vertex. \\
$|\Delta\phi(\vec{p}_{\mathrm{vtx}},\vec{p}_{\gamma})|$
& Azimuthal separation between the momentum of the tracks associated with the vertex and the photon direction. \\
$\dphimetjlead$
& Azimuthal separation between the missing transverse momentum and the leading jet. \\
$|\Delta\phi(\text{j}_1,\text{j}_2)|$
& Azimuthal separation between the two leading jets. \\
$\dmet$
& Difference between the missing transverse momentum computed without JVT requirements and the nominal reconstruction. \\
\metsig
& Significance of the missing transverse momentum. \\
$(z_{\mathrm{vtx}}-z_{\gamma})/\sigma(z_{\gamma})$
& Distance along the longitudinal ($z$) axis between the reconstructed primary vertex and the vertex position inferred from the photon trajectory (photon pointing), normalised by the photon pointing resolution. The photon trajectory is reconstructed using its energy deposits in the first and second layer of the ECAL. \\
$z_{\mathrm{skew}}$
& Skewness of the $z$ distribution of tracks associated with the vertex. \\
Number of central jets
& Number of reconstructed jets in the central region ($|\eta|<2.5$). \\
$(\pTmiss+\pTy)/\sum{p_{\mathrm{T}}^{\mathrm{jet}}}$
& Transverse momentum balance between the photon+$\ETmiss$ system and the reconstructed jets. \\
$z_{\mathrm{kurt}}$
& Kurtosis of the $z$ distribution of tracks associated with the vertex. \\
\bottomrule
\end{tabular}
\end{table}

\subsection{Signal region}
\label{sec:sel-SR}

\begin{table}[ht]
\centering
\small
\caption{Summary of the selections applied in the signal (SR) and control (CR) regions. In the CRs with muons, the \ETmiss\ is recomputed treating muons as invisible particles.}
\label{tab:region_summary}
\begin{tabular}{l|c|c|c|c}
\toprule
Selection & ~~SR~~ & ~~Low \metsig\ CR~~ & ~~$1\mu$ CR~~ & ~~$2\mu$ CR~~ \\
\midrule
Photon & \multicolumn{4}{c}{1 (\textit{selected}) with $\pt>50$~\GeV} \\
$\ETmiss$           & \multicolumn{4}{c}{$>100$~\GeV}  \\
$\mT$               & \multicolumn{4}{c}{$>80$~\GeV} \\
$\mathrm{BDT}_{\mathrm{vertex}}$ & \multicolumn{4}{c}{$>0.1$} \\
$N_{e,\tau}$        & \multicolumn{4}{c}{0 (\textit{baseline})} \\
$N_{b\textrm{-jets}}$ & \multicolumn{4}{c}{0} \\\midrule
$N_{\mathrm{jet}}^{\mathrm{central}}$ & \multicolumn{3}{c|}{$\leq 3$} & --- \\\midrule
$N_{\mu}$           & \multicolumn{2}{c|}{0 (\textit{baseline})} & 1 (\textit{selected}) & 2 (\textit{selected}) \\\midrule
$\metsig$ & $>6$ & $[2,6]$ & $>6$ & --- \\\midrule
$\dphimety$ & \multicolumn{4}{c}{$>1.25$} \\\midrule
$\dmet$             & \multicolumn{3}{c|}{$>-10$~\GeV} & --- \\
$\dphimetj$ & \multicolumn{3}{c|}{$<0.75$} & --- \\
$|\Delta\phi(j_1,j_2)|$ & \multicolumn{3}{c|}{$<2.5$} & --- \\
$|\etay|$ & \multicolumn{3}{c|}{$<1.75$} & --- \\
$m_{\mu\mu\gamma}$  & \multicolumn{3}{c|}{---} & $>100$~\GeV \\
\bottomrule
\end{tabular}
\end{table}

In addition to the \textit{baseline} selection,  the SR definition, summarised in Table~\ref{tab:region_summary}, includes requirements on the photon kinematics, angular separations between the photon, \ETmiss\ and jets, as well as further \ETmiss-related variables.
The latter include the \metsig\ and the difference $\dmet=\ETmissNoJVT-\ETmiss$, where $\ETmissNoJVT$ is defined in Section~\ref{sec:object-rec}.
In particular, the requirement of $\metsig>6$ further suppresses events with \textit{fake} \ETmiss, which typically exhibit lower significance values. Moreover, events are required to satisfy $\dmet > -10$~\GeV to reduce the contribution from incorrect PV identification. If a PU vertex is identified as the HS one, the standard \ETmiss\ computation rejects the HS jets while retaining PU jets. In this scenario, the $\ETmissNoJVT$ is expected to be closer to the true \ETmiss\ than the standard reconstruction, resulting in negative $\dmet$ values for events with no expected true \ETmiss.
Additionally, restricting the photon pseudorapidity to $|\eta_{\gamma}| < 1.75$ reduces background contributions from $W/Z+\gamma$ processes. Further requirements $\dphimety > 1.25$ and $\dphimetj < 0.75$ are imposed on the azimuthal separation between the \pTmissvec\ and its photon or jet term, $\vec{p}_{\mathrm{T}}^{\mathrm{miss},\gamma}$ and $\vec{p}_{\mathrm{T}}^{\mathrm{miss,jet}}$ (i.e. the negative vector sum of the $\vec{p}_\mathrm{T}$ of all photons or jets in the event). The selection on \dphimety\ is particularly effective in rejecting residual background from events with a misidentified PV while preserving the acceptance of the boosted Higgs boson signal. Finally the azimuthal separation between the two central jets with the highest \pT\ is required to be $|\Delta\phi(j_1, j_2)| < 2.5$.

The signal selection efficiency is 0.25\% for the ggF channel, and 1.06\%, 1.25\% and 0.66\% for the VBF, $WH$, and $ZH$ channels, respectively. The largest efficiency loss is due to the trigger decision, followed by the $\pTmiss$ selection, which retains 15\% of the events selected by the trigger.
The dominant Higgs boson production mode is ggF (68\% of the selected signal events), followed by VBF (23\%), $WH$ (6\%) and $ZH$ (3\%).

\begin{figure}[ht]
\centering
\includegraphics[width=0.95\linewidth]{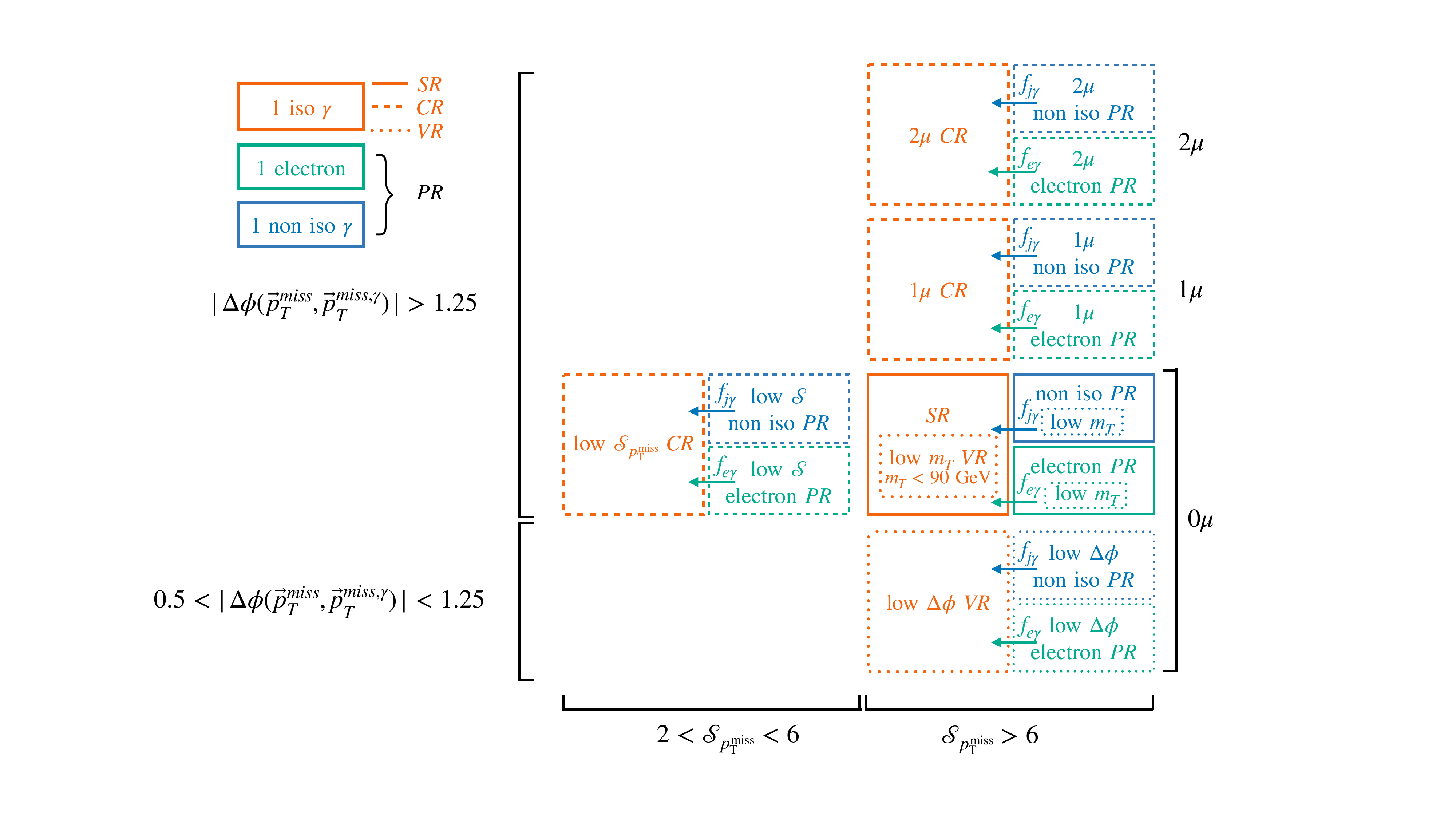}
\caption{Schematic overview of the signal (SR), control (CR),  validation (VR) and probe (PR) regions used in the analysis. Different colours correspond to different final states. Events with one isolated (iso) photon are selected in the SR (continuous lines), CRs (dashed lines) and VRs (dotted lines), while events with a single electron or non isolated photon are used for the corresponding PRs. Data in the PRs are scaled by the fake factors indicated by the arrows (\fey\ and \fjy) to estimate the \jfakey\ and \efakey\ contribution in each analysis region.}
\label{fig:regions_sketch}
\end{figure}

% End of text imported from the .//sections/selections.tex input file

\section{Analysis strategy and background modelling}
\label{sec:bkg}

% The next lines are included from the .//sections/background.tex input file
%

In addition to the SR, multiple orthogonal regions illustrated in Figure~\ref{fig:regions_sketch}, are used in the analysis. Three Control Regions (CRs), enriched in $W\gamma$, $Z\gamma$ and \jfakey\ contributions and defined in Table~\ref{tab:region_summary}, are used to constrain the corresponding backgrounds in a maximum-likelihood fit. Validation Regions (VRs), designed to be kinematically close to the SR, are used to validate the modelling of \jfakey\ and \efakey. In addition, dedicated regions, referred to as Probe Regions (PRs), are employed for the \jfakey\ and \efakey\ data-driven estimates as detailed later.

The $1\mu$~CR and the $2\mu$ CR are employed to normalise the MC simulations of $W\gamma$ and $Z\gamma$ respectively to data. They are defined in a similar way to the SR, but requiring either one or two selected muons. To emulate the SR kinematics, muons are treated as invisible in the \ETmiss\ reconstruction. In the $2\mu$ CR, some of the SR requirements, as detailed in Table~\ref{tab:region_summary} are removed to increase the available number of events. The contribution from the  \yjetdirect\ background due to fake \ETmiss\ is estimated purely from MC, as it accounts for about $2\%$ of the total.

The background from \efakey\ or \jfakey\ is estimated in all the analysis regions by rescaling data in PRs, by fake factors evaluated as described in Sections~\ref{sec:fey} and~\ref{sec:fjy}. For the \efakey\ background, electron PRs are defined for each analysis region, replacing the selected photon by a selected electron, and an electron trigger is also required, for consistency with the evaluation of the \efakey\ fake factors, as explained in Section~\ref{sec:fey}. For the \jfakey\ background, PRs with a non-isolated photon, \yNI, are introduced by requiring $\isol>0.1$, in order to ensure a negligible contamination from genuine photons.

\begin{figure}[htpb]
\centering
\subfloat[]{\includegraphics[trim=5cm 1cm 5cm 1cm, clip, width=0.45\linewidth]{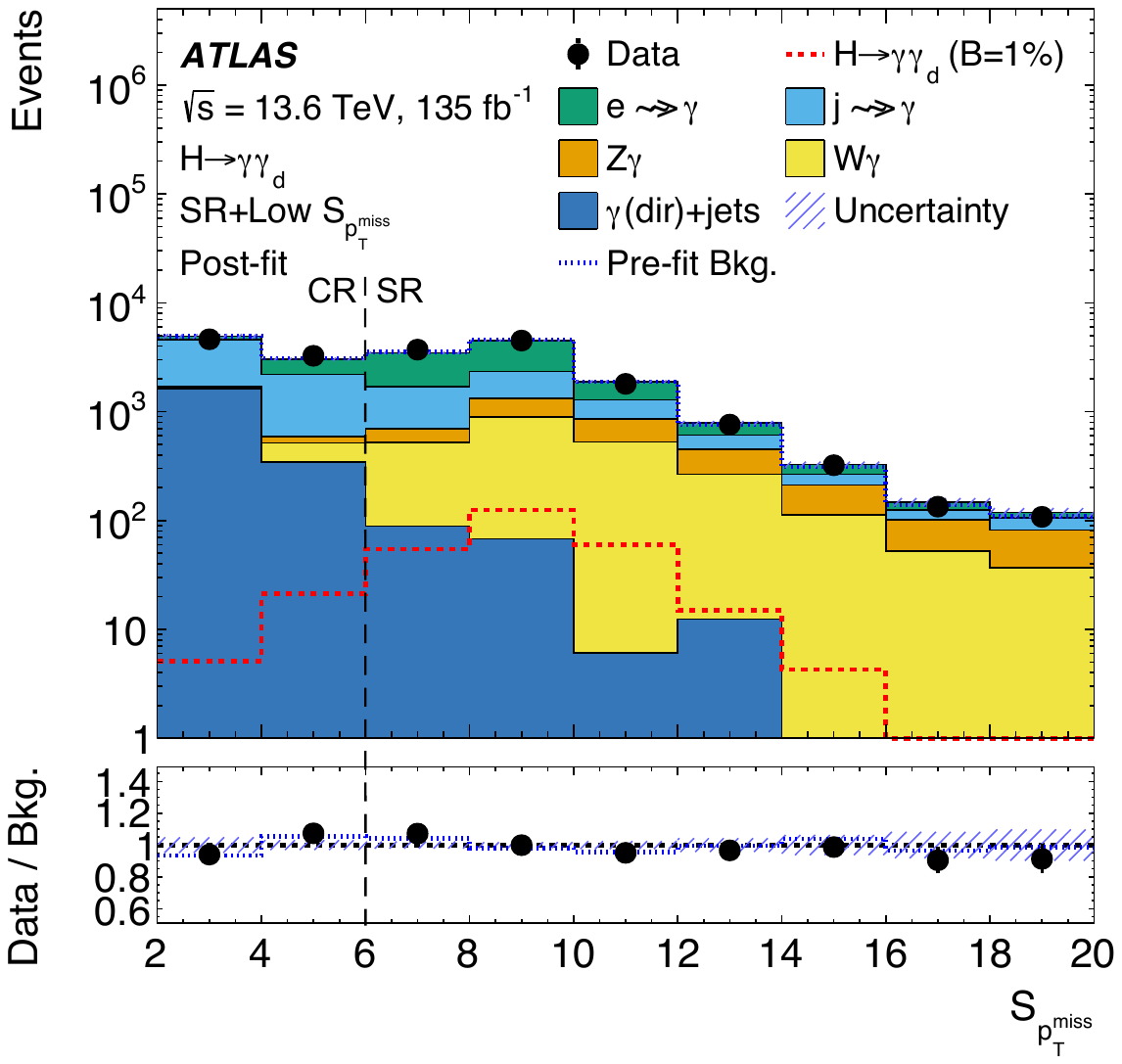}\label{fig:metsig_dist}}
\subfloat[]{\includegraphics[trim=5cm 1cm 5cm 1cm, clip, width=0.45\linewidth]{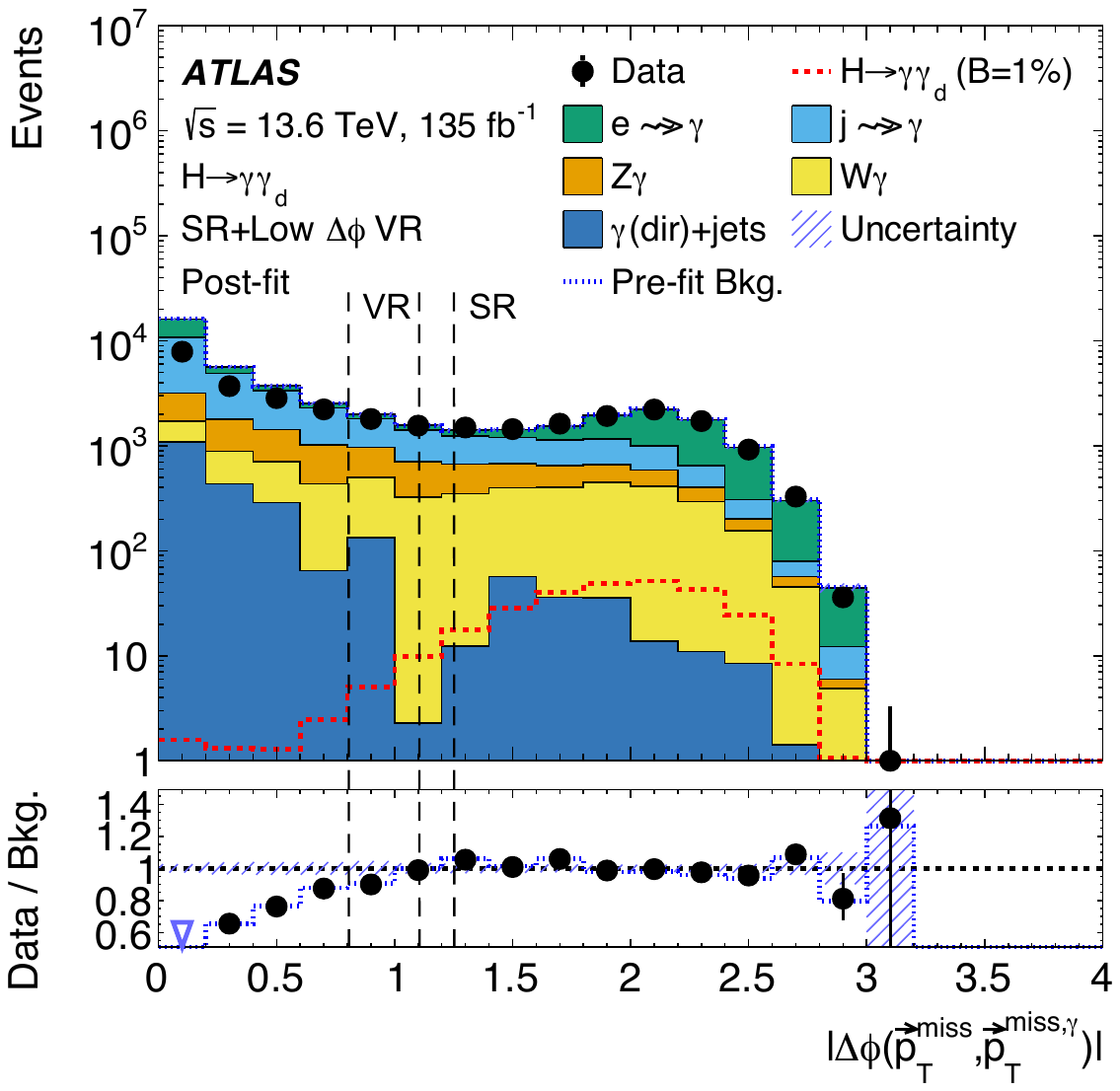}\label{fig:dphi_dist}}
\caption{Post-fit distributions of (a) \metsig\ in the signal region (SR) and low-\metsig\ control region (CR), and of (b) \dphimety\ in the SR and low-$\Delta\phi$ validation region (VR), for data and the expected SM background. The dashed vertical lines mark the boundaries of the SR and the CR/VR\@. The $H\to\gamma\gammad$ signal, assuming $\br=1\%$, is overlaid as a dotted line and includes the ggF, VBF, $WH$, and $ZH$ production modes. The pre-fit prediction is overlaid as a dashed line. The uncertainties include both the statistical and the systematic components. The lower panels show the ratios of data to expected background yields post-fit (dots) and pre-fit (dotted line). The downward-pointing triangle marks a ratio point located below the bottom edge of the ratio plot.
}
\end{figure}

Due to the large systematic uncertainty affecting the \jfakey\ component, the low~\metsig~CR, defined by requiring $2 <\metsig <6 $, is included in the fit to provide an additional constraint on this background. The post-fit \metsig\ distribution is shown in Figure~\ref{fig:metsig_dist}, with the dashed line highlighting the boundary between the low~\metsig~CR and the SR\@. The \efakey\ contribution is further constrained in the low \mT\ region of the SR that is dominated by $W(e\nu)+$jets events with the electron being misidentified as a photon, yielding a \mT\ value close to the $W$-boson mass. In particular, the low \mT\ VR, with $80<\mT<90$~\GeV, has very little signal, and is also used to validate the \efakey\ background before performing the fit. In addition, a low $\Delta\phi$ VR, with $0.8<\dphimety<1.1$, provides a sample orthogonal both to the SR and the low \metsig\ CR to also validate the modelling of the \jfakey\ background at high \metsig\ values. Figure~\ref{fig:dphi_dist} shows the post-fit \dphimety\ distribution for events passing all the SR requirements except for the \dphimety\ one, highlighting the SR and low $\Delta\phi$ VR boundaries with dashed vertical lines. This observable is particularly sensitive to PV misidentification at low values, which corresponds to events with low \pT\ or no jets, where the photon and the \met\ are nearly back-to-back. A residual mismodelling in this region is therefore expected even after applying the $\mathrm{BDT}_{\mathrm{vertex}}$ requirement, primarily due to events with a correct PV being misclassified by the BDT as an incorrect PV\@. Despite this, the agreement is relatively good and it improves with increasing \dphimety\ values, as expected.

\subsection{Estimation of \efakey}
\label{sec:fey}

The fake factor for \efakey\ estimation is defined as:
\begin{equation}
\fey=\frac{N(\efakey)}{N(e\rightsquigarrow e)},
\label{eq:def-fey}
\end{equation}
where $N(\efakey)$ and $N(e\rightsquigarrow e)$ denote respectively the numbers of electrons reconstructed as photons or as electrons.
The values of the \fey\ are computed directly from data, exploiting the $Z\to\ee$ process, where prompt isolated photons are absent.
While $N(\efakey)$ is counted from reconstructed photons in \ye\ events selected by a single electron trigger, $N(e\rightsquigarrow e)$ is counted from both $e^+$ and $e^-$ in \ee\ events selected by a dielectron trigger.
Both numbers are calculated in an invariant mass window $m_\text{inv}\in[80,100]~\GeV$ around the $Z$ peak, $m_\text{inv}$ being the invariant mass of the \ye\ or the \ee\ pair.

\begin{figure}
\centering
\includegraphics[width=0.6\textwidth]{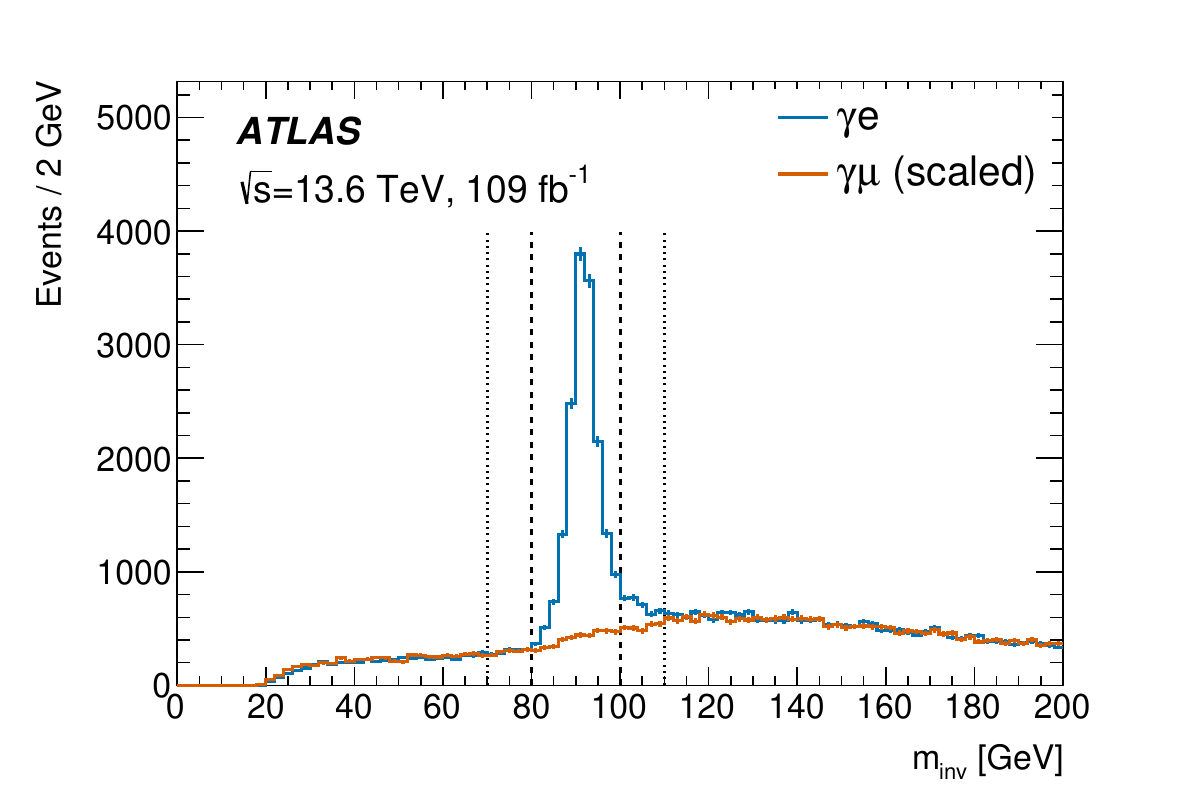}
\caption{Invariant mass of reconstructed $\gamma e$ and $\gamma\mu$ pairs from data, the latter being rescaled so that its sidebands match those of the former. The vertical dashed lines show the window used to compute \fey, while the vertical dotted lines mark the sidebands used for the normalisation of the $\gamma\mu$ distribution. Plots are inclusive in $\pTy$ and $\etay$, and show only data taken in 2024. Similar plots are obtained from 2023 data.}
\label{fig:efakey_minv}
\end{figure}

While the \ee\ events provide a rather pure sample of $Z\to\ee$, the \ye\ events contain a large number of genuine photons, mostly from the $W(\to e\nu)\gamma$ process, which must be subtracted. This is achieved by exploiting lepton universality, and selecting a sample of $\gamma\mu$ events. To account for the different reconstruction efficiency for $e$ and $\mu$, the invariant mass spectrum of the $\gamma\mu$ events is normalised to that of the $\gamma e$ in two sidebands $m_\text{inv}<70~\GeV$ and $m_\text{inv}>110~\GeV$, as displayed in Figure~\ref{fig:efakey_minv}.
The complete definition of the fake factor is therefore:
\begin{equation}
\fey=\dfrac{N_{\gamma e}^\mathrm{peak}-N_{\gamma\mu}^\mathrm{peak}\dfrac{N_{\gamma e}^\mathrm{side}}{N_{\gamma\mu}^\mathrm{side}}}{N_{ee}^\mathrm{peak}}.
\label{eq:calc-fey}
\end{equation}

The value of \fey\ is computed separately for 2023 and 2024 datasets as a function of \etay, since no significant dependence on \pTy\ is observed. The values of \fey\ are weakly dependent on the choice of the peak width and on the sidebands. Each variation is treated as a systematic uncertainty, and are all combined in quadrature with the statistical uncertainty. The values of \fey\ range between $\sim0.027$ in the barrel and $\sim0.054$ in the endcaps, with overall relative uncertainties between $3\%$ and $8\%$.

The $e$-PR must satisfy both the analysis trigger, to guarantee the same event topology as the SR and CRs, and the electron trigger, for consistency with the \fey\ evaluation.
However, the \fey\ evaluation does not consider the analysis trigger, therefore a further correction is necessary, to account for the efficiency of the analysis trigger on reconstructed $\gamma$ and $e$:
\begin{equation}
\fey^\mathrm{corr} = \fey \frac{\epsilon_\gamma^\mathrm{trig}}{\epsilon_e^\mathrm{trig}}.
\label{eq:calc-fey-corr}
\end{equation}
where $\epsilon_\gamma^\mathrm{trig}$ and $\epsilon_e^\mathrm{trig}$ are the fraction of events that pass the analysis trigger, in events containing a genuine photon (SR-like) and a genuine electron ($e$-PR-like) respectively.
The only trigger requirement affecting the evaluation of \fey\ is that on the reconstructed photon, therefore the effects of the \pTmiss\ and \mT\ selections need to be disentangled.
This is achieved by applying offline cuts on \pTmiss\ and \mT\ higher than the trigger-level cuts, to ensure being in the plateau of the trigger efficiency turn-on curves.
To evaluate $\epsilon_\gamma^\mathrm{trig}$, $\gamma\mu$ events were selected by a muon trigger. The muon is treated as invisible in the offline evaluation of \pTmiss, and $\ETmissInvMuon>150~\GeV$ and $(\mT)^{\mu\textrm{-inv}}>100~\GeV$ are required. The trigger-level evaluation of \pTmiss\ does not consider muons, therefore they are also treated as invisible by the analysis trigger. The fraction of events that satisfy the analysis trigger criteria provides the value of $\epsilon_\gamma^\mathrm{trig}$.
Similarly, $\epsilon_e^\mathrm{trig}$ is evaluated on events with an electron, selected from data by an electron trigger, and fulfilling $\pTmiss>200~\GeV$ and $\mT>100~\GeV$. The fraction of events that satisfy the analysis trigger criteria provides the value of $\epsilon_e^\mathrm{trig}$.
Both $\epsilon_\gamma^\mathrm{trig}$ and $\epsilon_e^\mathrm{trig}$ are computed as a function of \pT, to better track the trigger efficiency turn-on. While $\epsilon_\gamma^\mathrm{trig}$ reaches full efficiency for $\pT>60$~\GeV, $\epsilon_e^\mathrm{trig}$ saturates at about 90\%, as the photon trigger is not fully optimal for electrons.
The variations of
$\epsilon_\gamma^\mathrm{trig}$ and $\epsilon_e^\mathrm{trig}$ when applying different cuts on \pTmiss\ and \mT\ are treated as systematic uncertainties and combined in quadrature. The resulting trigger correction factor is between 1.06 and 1.26 with an overall relative uncertainty between $3\%$ and $12\%$.

\begin{figure}[htbp]
\centering
\includegraphics[width=0.45\textwidth]{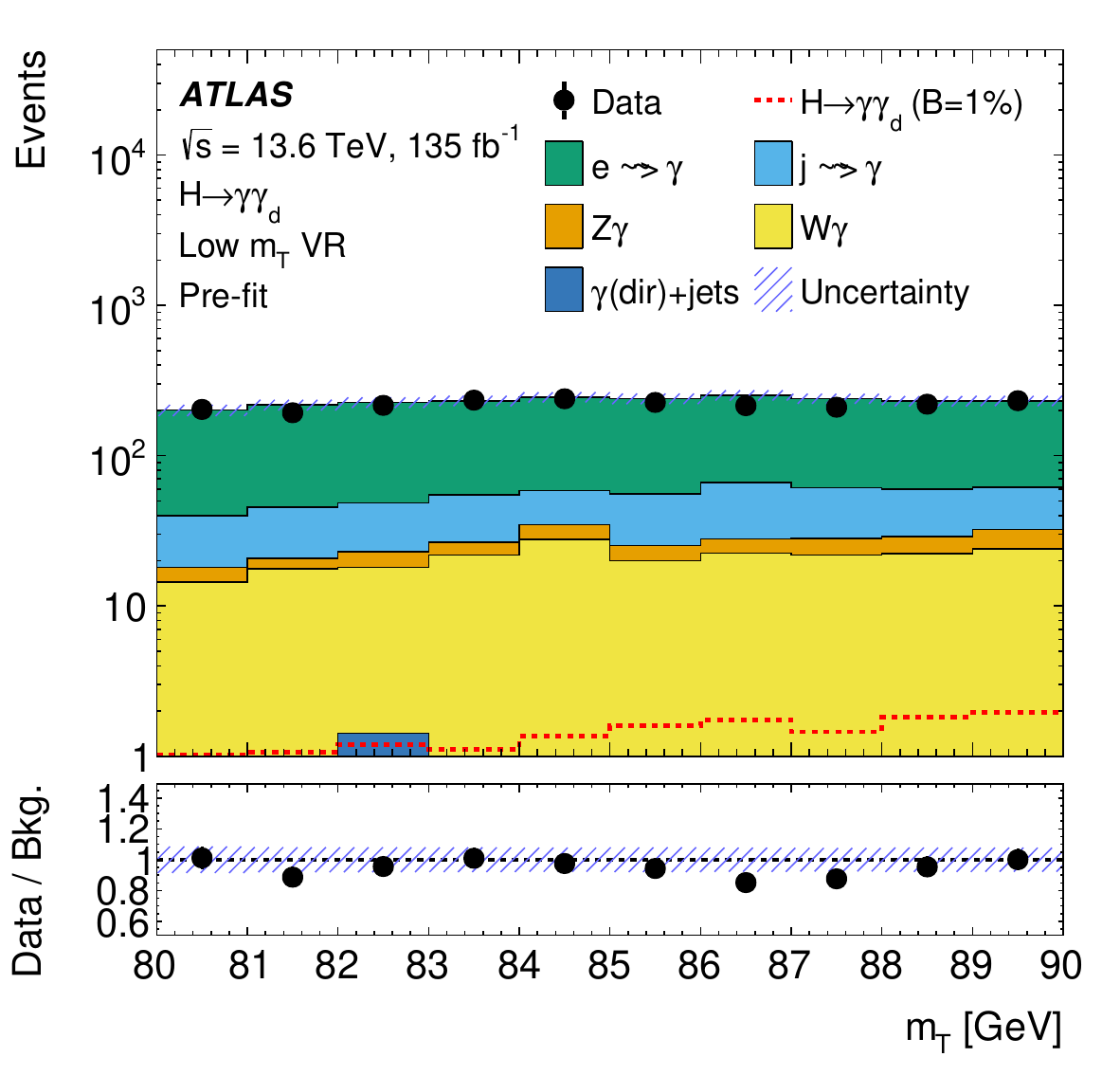}
\caption{Pre-fit distribution of \mT\ for data and the expected SM background in the low-\mT\ VR\@. The $H\to\gamma\gammad$ signal, assuming $\br=1\%$, is overlaid as a dashed line and includes the ggF, VBF, $WH$, and $ZH$ production modes. The uncertainties include both statistical and systematic components. The lower panel shows the ratio of data to expected background yields.}
\label{fig:efakey-VR}
\end{figure}

The validation of the \efakey\ background in the low \mT\ VR is displayed in Figure~\ref{fig:efakey-VR}, showing good agreement before the fit.

\subsection{Estimation of \jfakey}
\label{sec:fjy}

The \jfakey\ fake factor is defined as
\begin{equation}
\fjy=\frac{N(\jfakey)}{N(j\rightsquigarrow\yNI)}.
\label{eq:def-fjy}
\end{equation}
where $N(\jfakey)$ and $N(j\rightsquigarrow\yNI)$ denote the numbers of jets reconstructed as isolated or non-isolated photons, respectively.
Isolated and non-isolated reconstructed photons are defined by requiring isolation $\isol<0.022$ and $\isol>0.1$, respectively. The evaluation of the fake factor relies on the distribution of the isolation for \jfakey\ satisfying the \T\ identification, $\Phi^\T(\isol)$, and is calculated as:
\begin{equation}
\fjy=\frac{\int_{-\infty}^{+0.022}\text{d}\isol\,\Phi^\T(\isol)}{\int_{+0.1}^{+\infty}\text{d}\isol\,\Phi^\T(\isol)}.
\label{eq:calc-fjy}
\end{equation}

The $\Phi^\T(\isol)$ distribution cannot be directly extracted from data, as the \T\ photon selection contains comparable contributions from \jfakey\ and genuine photons. Moreover, $\Phi^\T(\isol)$ is not well modelled by the MC, and depends on the MC generator used. Therefore $\Phi^\T(\isol)$ has to be inferred, as explained in the following.
Photons satisfying the \Lp\ selection\footnote{
\Lp\ photons are selected in events satisfying the analysis trigger. This is possible, as the \T\ selection at trigger level is looser than at analysis level. The contamination of genuine photons in the \Lp\ sample is estimated to be $\lesssim5\%$, from simulations.
}
(see Section~\ref{sec:object-rec}) are enriched in \jfakey, therefore the isolation distribution $\Phi^\Lp(\isol)$ of \jfakey\ in the \Lp\ selection can be extracted from data. However, the photon selection affects the isolation, making $\Phi^\Lp$ broader than $\Phi^\T$. To infer $\Phi^\T$ from $\Phi^\Lp$, an affine transform is applied to the \isol\ of all \Lp\ photons:
\begin{equation}
\isol^\T = m^\T + \frac{w^\T}{w^\Lp}\left(\isol^\Lp-m^\Lp\right),
\label{eq:affine-transf}
\end{equation}
then all the obtained values of $\isol^\T$ are used to build $\Phi^\T$.
In Eq.~(\ref{eq:affine-transf}), $m$ and $w$ are the median and the 16\%-to-84\% half-width of the isolation distributions in the \Lp\ and \T\ samples. While $m^\Lp$ and $w^\Lp$ can be computed from the observed $\Phi^\Lp$ in data, $m^\T$ and $w^\T$ are inferred,
with the help of \jfakey\ from $W$+jets and $Z$+jets MC samples,
assuming that
\begin{equation}
\left(\frac{m^\T-m^\Lp}{w^\Lp}\right) = R_m \left(\frac{m^\T_{\MC}-m^\Lp_{\MC}}{w^\Lp_{\MC}}\right)
~,~~~~~~
\left(\frac{w^\T}{w^\Lp}\right) = R_w \left(\frac{w^\T_{\MC}}{w^\Lp_{\MC}}\right)
~,
\label{eq:def-Rm-Rw}
\end{equation}
where $m^\T_\MC,~m^\Lp_\MC,~w^\T_\MC,~w^\Lp_\MC$ are extracted from the $Z$+jets and $W$+jets MC samples.
The values of $R_m$ and $R_w$ are estimated by exploiting the various \Lpp\ photon selections (see Section~\ref{sec:object-rec}), which are closer to the \T\ selection.
The \Lpp\ selections in data are still dominated by \jfakey, therefore $m^\Lpp$ and $w^\Lpp$ (i.e. the median and half-width in the \Lpp\ selection) can be evaluated in data from the observed $\Phi^\Lpp$, and Eq.~(\ref{eq:def-Rm-Rw}) can be used with each of the \Lpp\ instead of the \T\ selection, to derive $R_m$ and $R_w$. Different choices of \Lpp\ selections (labelled $\Lpp_\mathrm{a}$ in the following, where $a$ flags the different definitions) provide different values of $R_m$ and $R_w$, and therefore different $\Phi^\T_\mathrm{a}(\isol)$.
Since each of the \Lpp\ selections is obtained by applying at most one requirement on one of the strips shower shapes, this procedure allows the effect of one strips shower shape at a time to be probed in the $\Lp\to\T$ extrapolation.

\begin{figure}[htbp]
\centering
\includegraphics[width=0.6\textwidth]{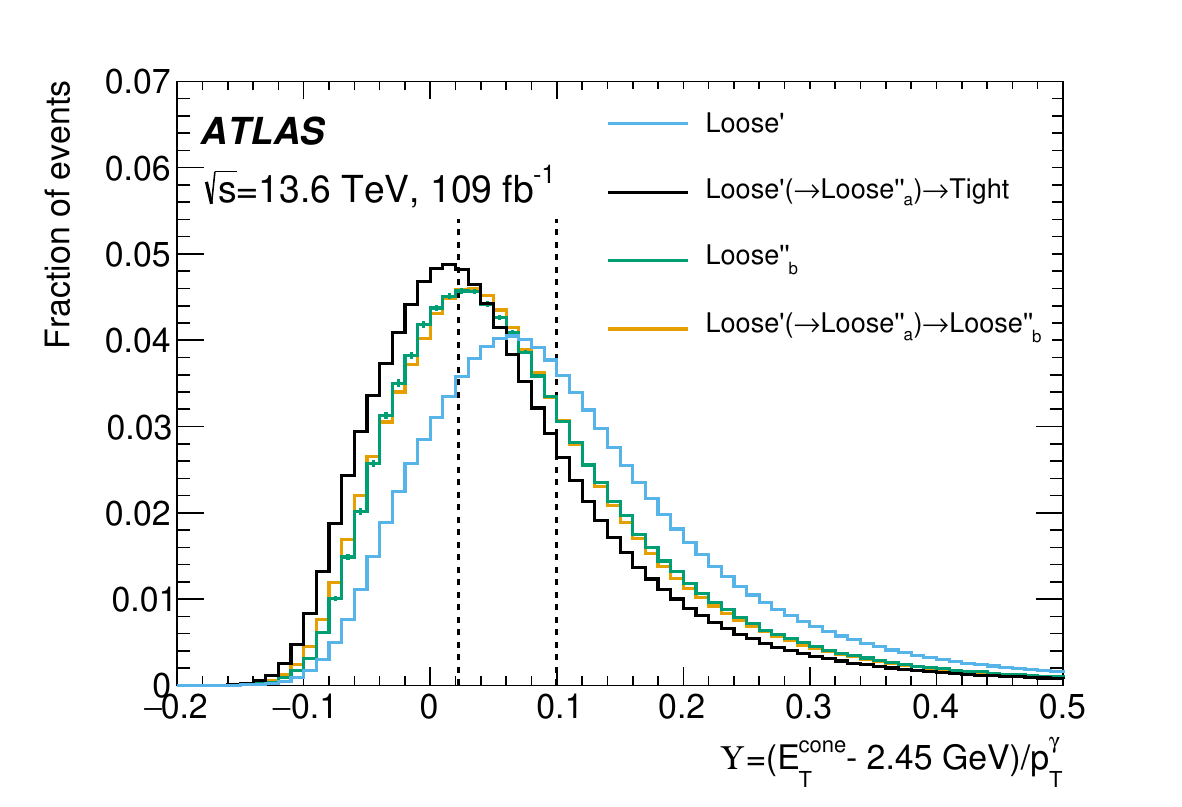}
\caption{Distributions of the isolation $\Upsilon=(E_T^\mathrm{cone}-2.45~\mathrm{\GeV})/p_T^\gamma$ for reconstructed photons, used for the \jfakey\ background estimate. The figure displays the distribution from the \Lp\ photon sample (enriched in \jfakey), and an example of extrapolated distribution for the \T\ sample ($\Lp(\to\Lpp_\mathrm{a})\to\T$), using a $\Lpp_\mathrm{a}$ sample to compute $R_m$ and $R_w$.
The two dashed vertical lines mark the `isolated' ($\isol<0.022$) and the `non-isolated' ($\isol>0.1$) regions.
Plots are inclusive in $\pTy$ and $\etay$, and show only data taken in 2024. Similar plots are obtained from 2023 data.
The figure also shows the observed and extrapolated distributions for a different $\Lpp_\mathrm{b}$ sample.}
\label{fig:jfakey_isol_extr}
\end{figure}

In summary, the affine transform of Eq.~(\ref{eq:affine-transf}) depends on the $R_m$ and $R_w$ parameters. These are evaluated to achieve a $\Lp\to\Lpp_\mathrm{a}$ extrapolation that reproduces at best the observed $\Lpp_\mathrm{a}$ sample, and are then applied to the $\Lp\to\T$ extrapolation. This process is symbolised as \mbox{$\Lp(\to\Lpp_\mathrm{a})\to\T$}, and provides an extrapolated distribution $\Phi^\T_\mathrm{a}(\isol)$. The observed isolation distribution for \Lp\ photons and one of the extrapolated distributions for \T\ photons are displayed in Figure~\ref{fig:jfakey_isol_extr}.
The reliability of the extrapolation is validated by comparing several extrapolated distributions $\Lp(\to\Lpp_\mathrm{a})\to\Lpp_\mathrm{b}$ with the observed $\Lpp_\mathrm{b}$ ones. An example is shown in Figure~\ref{fig:jfakey_isol_extr}.

For each extrapolated distribution $\Phi^\T_\mathrm{a}(\isol)$, a different fake factor $(\fjy)_\mathrm{a}$ is computed. The final fake factor is computed as the arithmetic mean of all $(\fjy)_\mathrm{a}$, and a conservative systematic uncertainty is set as half the envelope of all their values.

The fake factor is computed separately for 2023 and 2024 datasets as a function of \etay. The limited number of MC events prohibits studying its dependence on \pTy.
The values of \fjy\ range from $\sim1.5$ in the ECAL barrel, up to $\sim2.5$ in the endcaps, with relative uncertainties between 10\% and 15\%.

\begin{figure}[htbp]
\centering
\subfloat[]{\includegraphics[width=0.45\textwidth]{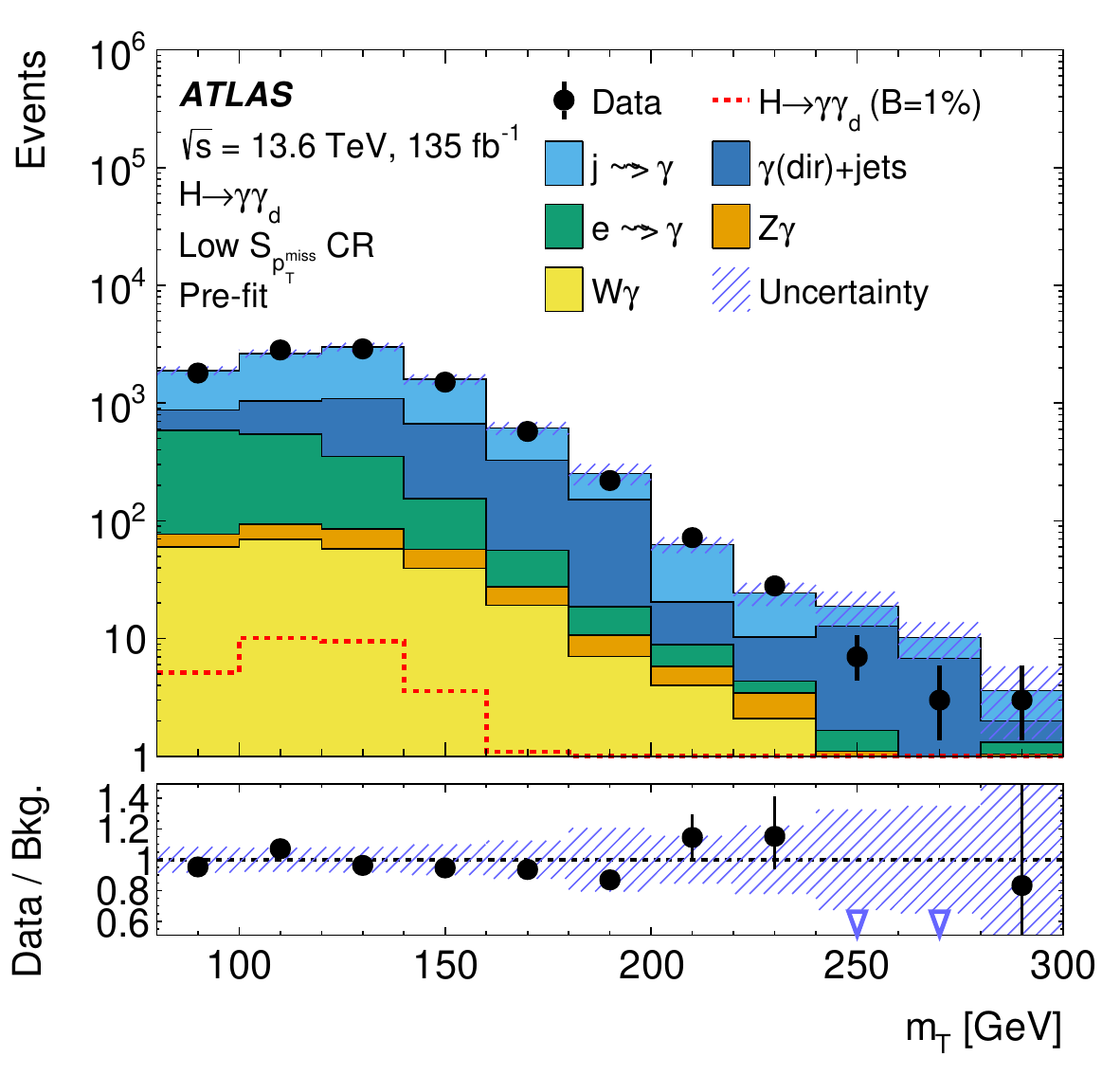}\label{fig:lowmetsig-vr}}
\hspace{0.05\textwidth}
\subfloat[]{\includegraphics[width=0.45\textwidth]{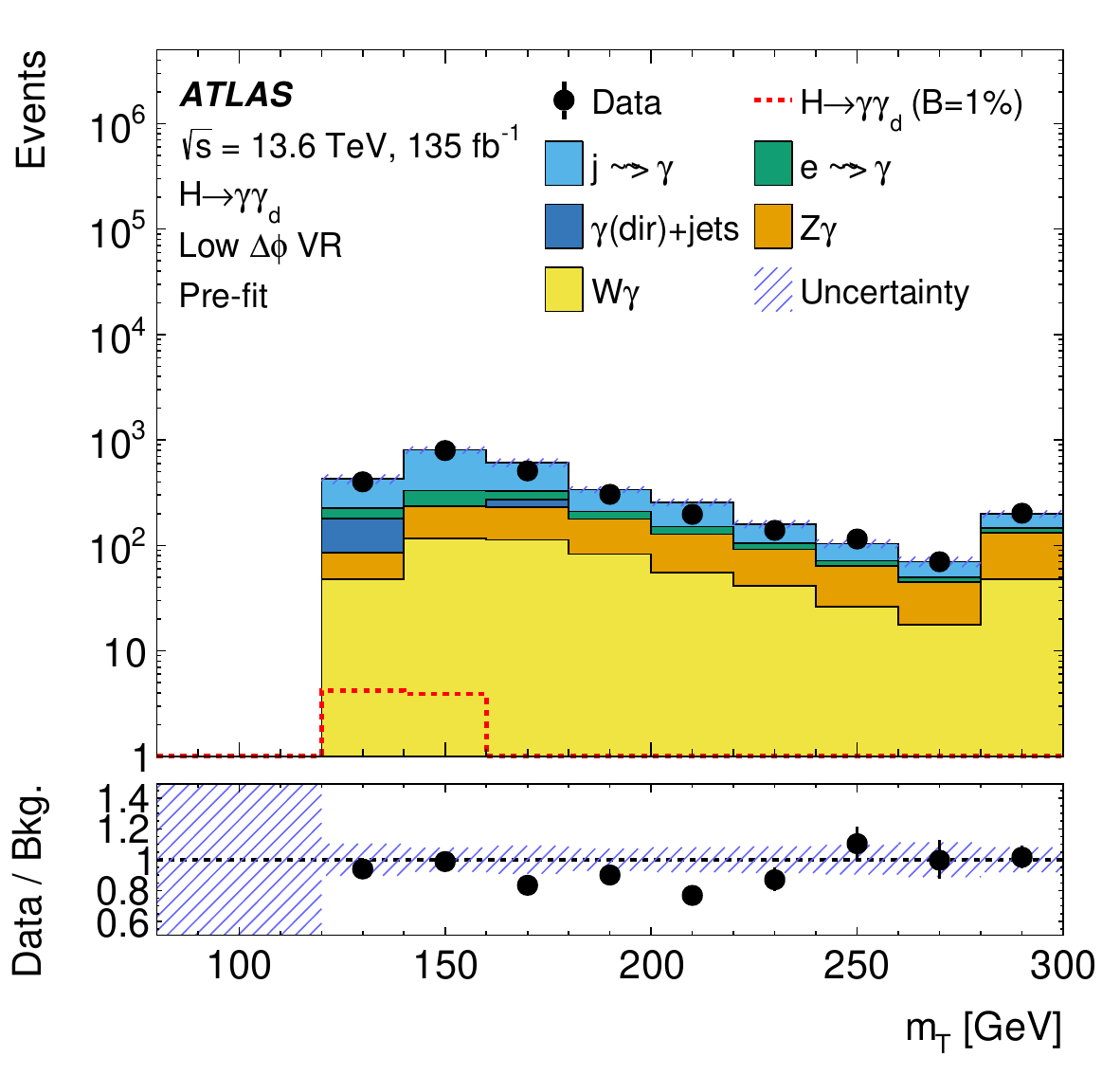}\label{fig:lowdphi-vr}}
\caption{Pre-fit distribution of \mT\ for data and the expected SM background in (a) the low \metsig\ CR and (b) the low $\Delta\phi$ VR, enriched in \jfakey. The $H\to\gamma\gammad$ signal, assuming $\br=1\%$, is overlaid as a dashed line and includes the ggF, VBF, $WH$, and $ZH$ production modes. The uncertainties include both statistical and systematic components. The lower panel shows the ratio of data to expected background yields. The downward-pointing triangles mark ratio points located below the bottom edge of the ratio plot.}
\label{fig:VRs}
\end{figure}

The obtained \jfakey\ prediction is validated, before the fit, in both the low \metsig\ CR and the low \dphi\ VR, as shown in Figure~\ref{fig:VRs}.

% End of text imported from the .//sections/background.tex input file

\section{Systematic uncertainties}
%

% The next lines are included from the .//sections/systematics.tex input file
%

Multiple sources of systematic uncertainty are considered in the analysis. The experimental and theoretical uncertainties are applied only to the MC simulations of the signal and $V\gamma$ background, while their impact on the \yjetdirect\ simulation is expected to be negligible compared to its statistical fluctuations. Additional systematic uncertainties arise from the data-driven background estimates.
The impact of the different sources of uncertainty in the fitted value of \br\ is summarised in Table~\ref{tab:impacts}. The contributions are evaluated using the covariance matrix decomposition method~\cite{pinto2024} and include both statistical and systematic components. The dominant contribution to the total uncertainty arises from the statistical component (88\%), including contributions from both the data and the background. The dominant systematic contribution arises from experimental uncertainties (42\%), primarily driven by jet-related effects. This is followed by uncertainties associated with the $\efakey$ and $\jfakey$ data-driven methods (17\%), and by theoretical uncertainties (17\%).

\begin{table}[!htbp]
\center
\small
\caption{Post-fit relative impact of each source of uncertainty on the total uncertainty of the signal yields, as obtained through the covariance matrix decomposition method. The overall impacts of statistical, experimental, data-driven and theoretical systematic uncertainties are shown, followed by the individual contributions.}
\label{tab:impacts}
\begin{tabular}{ l  c  }
\toprule
\textbf{Category} & \textbf{Contribution to uncertainty in the \br} \\\midrule
\textbf{Statistical uncertainties} & 88\%\\
Data & 74\%\\
Background & 48\%\\\midrule
\textbf{Experimental systematic uncertainties} &  42\%\\
Luminosity & 4.6\%\\
Pile-up reweighting & 7\%\\
Pile-up corrections in jets calibration & 36\%\\
Jets & 18\%\\
Photons & 3.7\%\\
Muons & 2.1\% \\
\ETmiss & <1\% \\
Trigger & <1\% \\\midrule
\textbf{Systematic on data-driven methods uncertainties} &  17\%\\
$\jfakey$ & 11\%\\
$\efakey$ & 13\%\\
\midrule
\textbf{Theoretical systematic uncertainties} & 17\% \\\bottomrule
\end{tabular}
\end{table}

\subsection{Experimental uncertainties}

The uncertainty in the integrated luminosity is 1.9\%, derived from the calibration of the LUCID-2 detector and complemented by information from the ID and the calorimeters ~\cite{ATL-DAPR-PUB-2025-001}. This uncertainty is applied as a global normalisation uncertainty to all MC samples, including both signal and background processes. A systematic uncertainty is associated with the pileup reweighting procedure applied to simulated events to correct for mismodelling of PU interactions in simulation.

Further uncertainties are associated with the reconstruction and isolation efficiency and with the energy calibration and resolution of photons, muons and jets.   Photon efficiency uncertainties are estimated by using Run~2 data and MC simulations~\cite{EGAM-2018-01}, with an additional uncertainty included to account for differences between Run~2 and Run~3, as derived from simulation. The uncertainties related to photon energy calibration and resolution are evaluated following the methodology described in Ref.~\cite{PERF-2017-03}, using the Run~3 data collected between 2022 and 2024. All contributions are summed in quadrature and considered fully correlated in photon $\eta$, resulting in one systematic uncertainty associated with the energy calibration, and one with the energy resolution.

Multiple sources of uncertainties are considered for muons. The uncertainties affecting the muon momentum scale and resolution are separated into charge-dependent and charge-independent components~\cite{MUON-2018-03}. %
Additional uncertainties are associated with muon reconstruction and isolation efficiencies, as well as with the track-to-vertex association efficiency, each of which is split into statistical and systematic components~\cite{PERF-2015-10}.

Jet reconstruction is affected by uncertainties related to jet vertex tagging and to the jet energy scale (JES) and resolution (JER). The uncertainty in the NN-JVT and fJVT calibrations~\cite{PERF-2014-03,PERF-2016-06} is derived from $Z\to\mu\mu$ and $Z\to ee$ events in 2022--2023 data and is dominated by PU modelling uncertainties,  particularly at low \pT.
The JES and JER uncertainties~\cite{PERF-2014-07,PERF-2015-09,JETM-2018-05} include contributions from area subtraction, PU residual corrections, absolute energy scale, and global sequential calibration \cite{JETM-2022-01}, and are evaluated using Run~3 MC simulations. The uncertainty in the in situ calibration is derived from Run~2 data, with an additional uncertainty included to cover differences between Run~2 and Run~3. Further uncertainties account for residual effects related to variations in the vertexing algorithm, calorimeter noise thresholds, and modelling differences in the forward calorimeter region. Among these systematic uncertainties, the dominant contribution arises from PU residual corrections, which is quoted separately in Table~\ref{tab:impacts}. An additional jet-related source of systematic uncertainty arises from flavour tagging~\cite{FTAG-2018-01}.

The \ETmiss\ is affected by three sources of systematic uncertainty~\cite{JETM-2020-03,Balunas:2023fpu}, related to the modelling of its soft term and evaluated using Run~3 data collected in 2022--2023 and the corresponding MC simulations. One scale uncertainty is obtained by scaling up and down the parallel projection of the soft-term along the hard-term, while two resolution uncertainties are derived by smearing the parallel and perpendicular projections of the soft-term along the hard-term.

Finally, a systematic uncertainty is assigned to the trigger efficiency, as described in Section~\ref{sec:trigger}, to account for the observed discrepancy between data and MC simulations in the trigger efficiency curves.

\subsection{Uncertainties in data-driven methods}
Three additional sources of systematic uncertainty are associated with the data-driven estimates of the \efakey\ and \jfakey\ backgrounds.
As described in Section~\ref{sec:fey}, for the \efakey\ background, one systematic uncertainty accounts for the dependence of the $f_{e\gamma}$ fake factor on the peak width and the definition of the sidebands, while the second contribution is associated with the trigger correction factors.
The third data-driven systematic uncertainty is associated with the extrapolation from $\Phi^\Lp$ to $\Phi^\T$ in the \jfakey\ estimation,
as described in Section~\ref{sec:fjy}.

\subsection{Theoretical uncertainties}
Theoretical uncertainties related to the modelling of quantum chromodynamics (QCD) effects are considered for background and signal MC samples, arising from variations of the renormalisation and factorisation scales, the value of the strong coupling constant, and the PDF uncertainties. The latter are evaluated following the PDF4LHC21 prescription~\cite{Ball:2022oua}. The uncertainty associated with missing higher-order corrections is evaluated by independently varying the renormalisation and factorisation scales by factors of 0.5 and 2.0 relative to their nominal values and taking the envelope of the resulting variations, while the uncertainty related to the strong coupling constant $\alphas$ is obtained by varying its value by $\pm 0.001$.
%

% End of text imported from the .//sections/systematics.tex input file

\section{Statistical treatment and results}
%

% The next lines are included from the .//sections/statistics.tex input file
%

To maximise the sensitivity to the Higgs boson mass peak, the transverse mass \mT\ is used as the discriminating variable in the binned maximum-likelihood fit in all regions, except for the $2\mu$ CR, which is treated as a single inclusive bin due to limited statistical precision.
The \mT\ binning is optimised such that the bulk of the expected signal is contained in the two central bins, defined as $110<m_\text{T}<140$~\GeV and $140<m_\text{T}<200$~\GeV, while the lower and upper bins provide additional constraints on the background normalisation. In particular, the first bin ($80<m_\text{T}<110$~\GeV) contains the low-\mT\ VR and is dominated by the \efakey\ background, allowing an effective constraint on this contribution. This choice optimises the sensitivity by balancing the shape information against the statistical precision in each bin.

The likelihood function is constructed as the product of Poisson probability distribution functions describing the event counts in each \mT\ bin of the SR and CRs, as functions of the expected signal and background yields. The branching ratio \br\ is included in the fit as the parameter of interest (POI). In addition, two  normalisation factors, $k_{W\gamma}$ and $k_{Z\gamma}$, are applied to the $W\gamma$ and $Z\gamma$ background processes respectively, and included as free parameters in the fit, primarily constrained by the corresponding 1$\mu$ and 2$\mu$ CRs.
Systematic uncertainties affecting the signal and background yields are incorporated into the likelihood model as independent nuisance parameters with a Gaussian constraint centred at zero with a width corresponding to the associated uncertainty. The experimental and data-driven systematic errors are treated as fully correlated across all regions, and experimental systematic errors are also correlated amongst all MC samples. Theoretical uncertainties are treated as uncorrelated across samples. All nuisance parameters are assumed to be mutually uncorrelated. The fit model is implemented in the \textsc{TRExFitter}~\cite{TRExFitter} framework.

\begin{figure}
\centering
\subfloat[]{\includegraphics[width=0.4\linewidth]{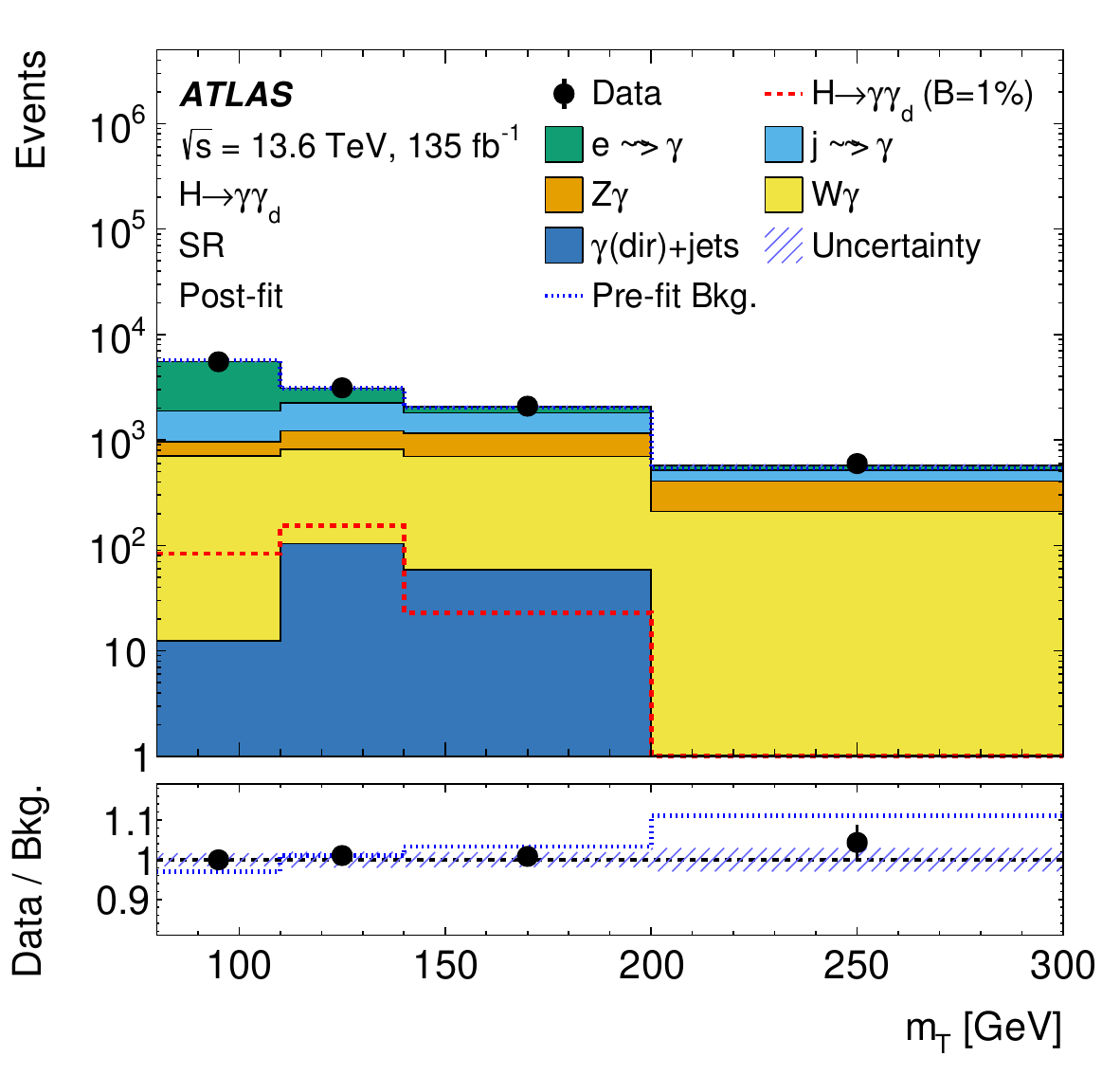}\label{fig:sr}}
\\
\subfloat[]{\includegraphics[width=0.4\linewidth]{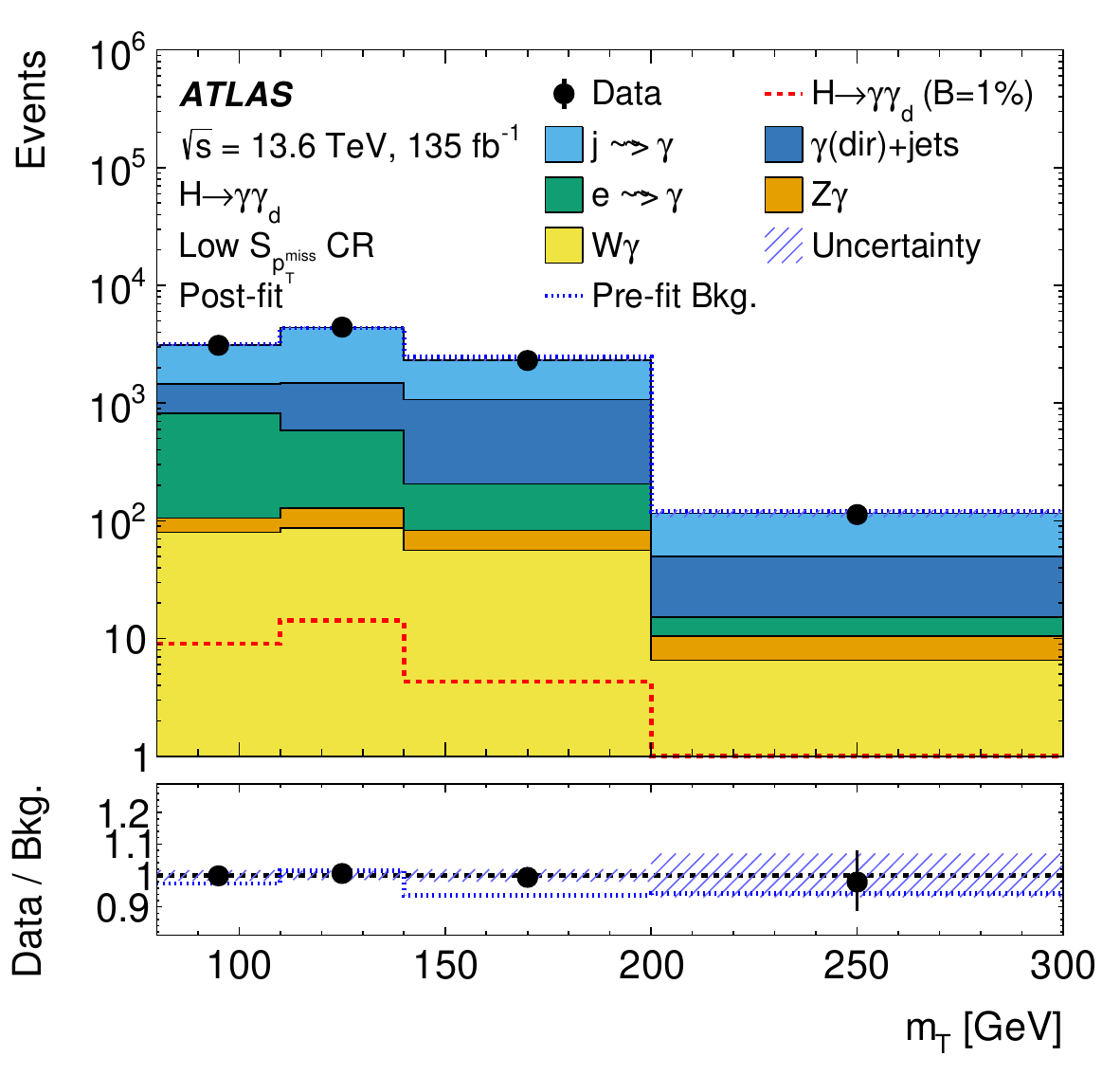}\label{fig:lowmetsig}}
\subfloat[]{\includegraphics[width=0.4\linewidth]{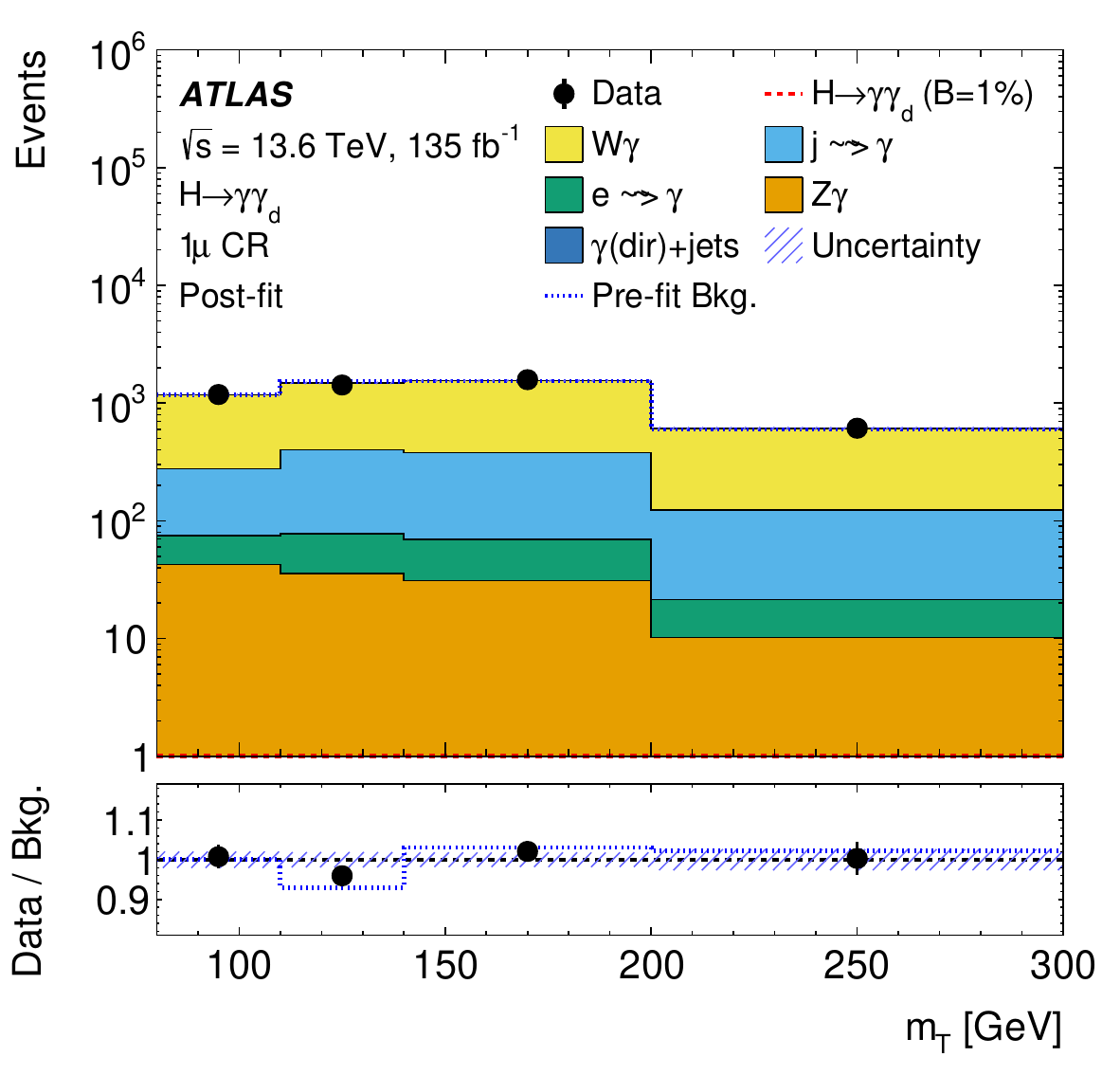}\label{fig:onemu}}
\caption{Distribution of \mT\ for data and for the expected SM background after the background-only fit, in the (a) SR and (b, c) CRs. The $H\to\gamma\gammad$ signal, assuming $\br=1\%$, is shown as a dashed line and includes the ggF, VBF, $WH$ and $ZH$ production modes. The pre-fit prediction is overlaid as a dotted line. The uncertainties include both statistical and systematic components determined by the multiple-bin fit. The lower panels show the ratios of data to expected background yields post-fit (dots) and pre-fit (dotted line).}
\label{fig:postfit_mt}
\end{figure}

\begin{figure}
\centering
\includegraphics[width=0.75\linewidth]{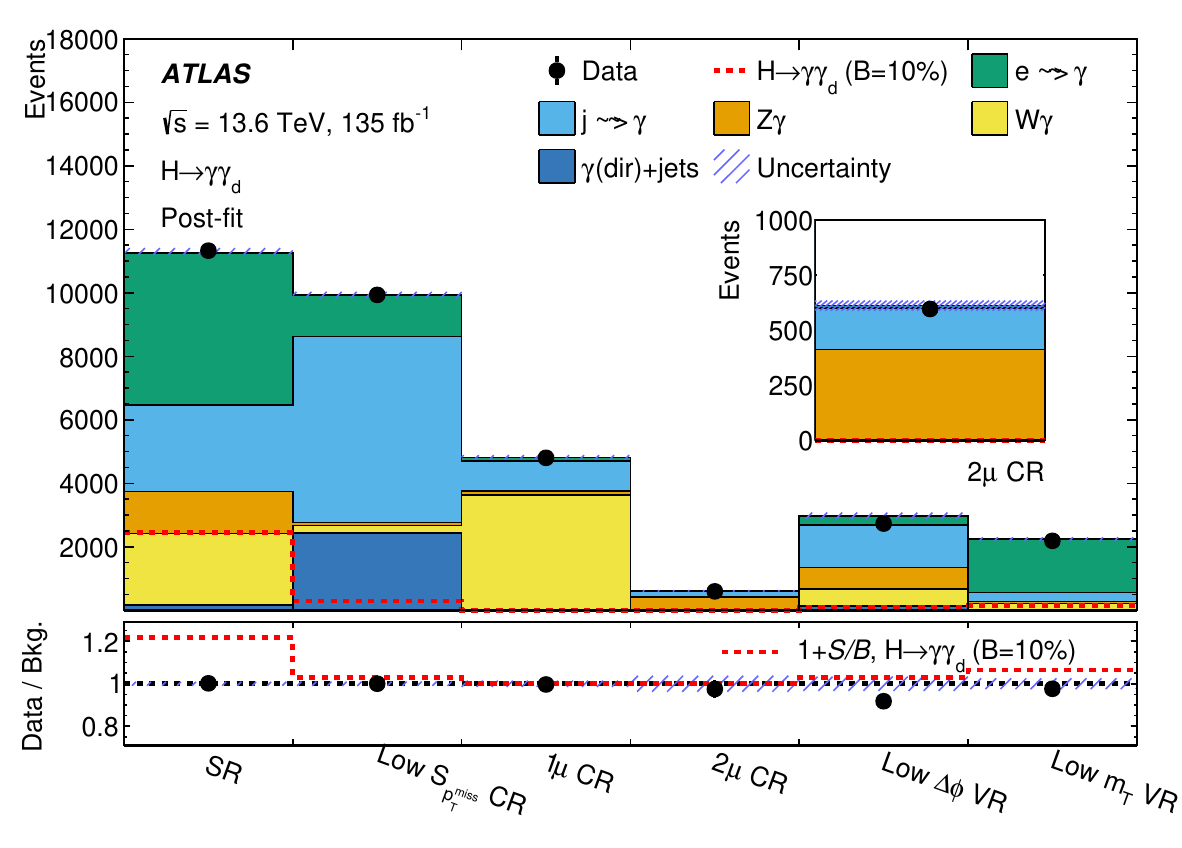}
\caption{
Observed data and post-fit background yields in the SR, CRs, and VRs after the background-only fit. The expected signal yields are overlaid, assuming $\br=10\%$ for visualisation purposes.
The lower panel shows the ratio of the data to the background, together with the signal contribution defined as $1 + S/B$, where $S$ and $B$ denote the signal and background yields.
}
\label{fig:summary_plot}
\end{figure}

A background-only fit is first performed in the SR and CRs, with \br\ set to zero, and  the $k_{W\gamma}$ and $k_{Z\gamma}$ normalisation factors left free. The fitted values are
$k_{W\gamma} = 0.93^{+0.08}_{-0.08}$ and
$k_{Z\gamma} = 1.03^{+0.12}_{-0.11}$.
The post-fit \mT\ distributions for data and the expected SM background are shown in Figure~\ref{fig:postfit_mt}, while the corresponding yields are reported in Table~\ref{tab:postfit_yields}. The yields and background composition in each analysis region, including the VRs, are also summarised in Figure~\ref{fig:summary_plot}. Good agreement is observed between data and the SM background expectation in all the analysis regions, and the data are consistent with the background-only hypothesis with an observed $p_0$-value of 0.37.

\begin{table}[htbp]
\centering
\small
\caption{Data and background yields in the SR and CRs, after the background-only fit. The pre-fit expected signal yields are also shown for each Higgs boson production channel, assuming $\br=1\%$. Contributions below $10^{-5}$ are shown as ``--''. Uncertainties include statistical and systematic components. The individual background uncertainties can be correlated and therefore do not add in quadrature to the total background uncertainty.}\label{tab:postfit_yields}
\begin{tabular}{
l
r@{\,$\pm$\,}l
r@{\,$\pm$\,}l
r@{\,$\pm$\,}l
r@{\,$\pm$\,}l
}
\toprule
&
\multicolumn{2}{c}{SR} &
\multicolumn{2}{c}{\lowMETsig\ CR} &
\multicolumn{2}{c}{1 $\mu$ CR} &
\multicolumn{2}{c}{2 $\mu$ CR} \\
\midrule

Pre-fit ggF $H \to \gamma\gammad$
& 166 & 24
& 18.2 & 4.9
& \multicolumn{2}{c}{--}
& \multicolumn{2}{c}{--} \\

Pre-fit VBF $H \to \gamma\gammad$
& 56.9 & 5.7
& 9.6 & 1.6
& \multicolumn{2}{c}{--}
& \multicolumn{2}{c}{--} \\

Pre-fit $WH$ $H \to \gamma\gammad$
& 14.7 & 1.4
& 1.41 & 0.31
& 0.149 & 0.018
& \multicolumn{2}{c}{--} \\

Pre-fit $ZH$ $H \to \gamma\gammad$
& 8.19 & 0.85
& 0.72 & 0.19
& 0.0073 & 0.0075
& 0.0249 & 0.0097 \\

\midrule

$Z\gamma$
& 1332 & 90
& 95 & 21
& 120 & 8.8
& 412 & 25 \\

$W\gamma$
& 2253 & 68
& 224 & 55
& 3635 & 80
& 0.677 & 0.058 \\

\efakey
& 4810 & 113
& 1302 & 43
& 124.5 & 3.5
& 11.46 & 0.45 \\

\jfakey
& 2726 & 105
& 5879 & 149
& 941 & 37
& 188 & 10 \\

\yjetdirect
& 175.7 & 3.6
& 2437 & 62
& 0.0469 & 0.0010
& \multicolumn{2}{c}{--} \\

\midrule

Total
& 11298 & 101
& 9936 & 98
& 4821 & 69
& 612 & 23 \\

\midrule

\multicolumn{1}{l}{Data}
& \multicolumn{2}{c}{11327}
& \multicolumn{2}{c}{9934}
& \multicolumn{2}{c}{4802}
& \multicolumn{2}{c}{596} \\

\bottomrule
\end{tabular}
\end{table}

\begin{figure}
\centering
\includegraphics[width=0.7\linewidth]{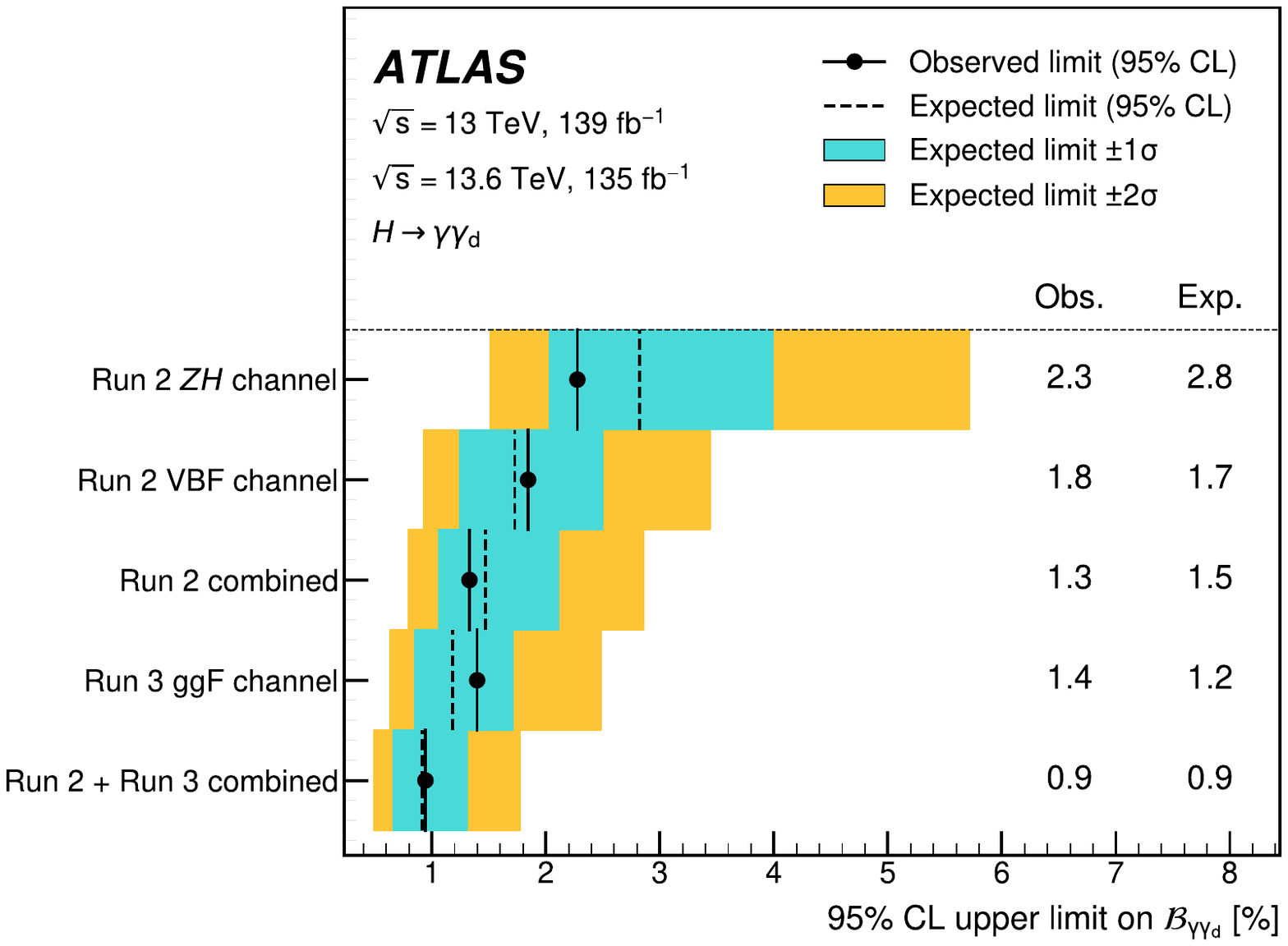}
\caption{Summary of the 95\%~CL upper limits on \br, achieved by the Run~2 searches (top three rows: $ZH$~\cite{HDBS-2019-13}, VBF~\cite{EXOT-2021-17}, and their combination~\cite{EXOT-2023-01}) and by this analysis (4th row). The overall Run~2~+~Run~3 combination is displayed in the bottom row.}
\label{fig:summary-limits}
\end{figure}

A second fit is then performed including the signal contribution and the branching ratio \br\ as the POI\@. The results are consistent with the background-only hypothesis, with
the best-fit value
$\br = 0.002^{+0.006}_{-0.006}$.
Since no significant excess over the SM prediction is observed, upper limits on the branching ratio of the exotic decay $H\to\gamma\gammad$ are set using the profile likelihood ratio test statistic and the CL$_\text{s}$ approach~\cite{Read:2002hq}, in the asymptotic approximation~\cite{Cowan:2010js}. An observed (expected) upper limit on \br\ of
$1.4\%$ ($1.2\%$) is obtained at 95\%~CL\@.
This result can be compared with previous searches carried out on the VBF~\cite{EXOT-2021-17} and $ZH$~\cite{HDBS-2019-13} production channels in Run~2, as well as their combination ~\cite{EXOT-2023-01}, as summarised in Figure ~\ref{fig:summary-limits}. The present analysis provides constraints on the $H\to\gamma\gammad$ decay comparable to the observed (expected) upper limits of 1.3\% (1.5\%) from the Run~2 combination, which are the most stringent limits from previous searches.
A combination of this analysis with Run~2 results is performed, treating systematic uncertainties as uncorrelated between Run~2 and Run~3 searches, and adopting for the different Run~2 channels the same correlation scheme as in Ref.~\cite{EXOT-2023-01}. As a conservative approach, the signal modelling is treated as uncorrelated between Run~2 and Run~3 searches, as the former were not sensitive to the ggF production mode, which is the dominant contribution in the present analysis, while the VBF and $VH$ channels account for only $\sim 30\%$ of the signal yield.
The combination yields an observed (expected) upper limit of 0.9\% (0.9\%) on \br.

% End of text imported from the .//sections/statistics.tex input file

\FloatBarrier

\section{Conclusion}
%

% The next lines are included from the .//sections/conclusion.tex input file
%

A search for semi-visible decays of the SM Higgs boson into a photon and \ETmiss\ is presented. The analysis exploits a dedicated topological trigger with reduced \ETmiss\ and \pTy\ thresholds, allowing the gluon--gluon fusion production channel to be targeted for the first time in ATLAS and extending the sensitivity to a previously unexplored kinematic region.
The analysis is performed using Run~3 $pp$ collision data at $\sqrt{s}=13.6$~\TeV, collected by the ATLAS detector in 2023 and 2024, corresponding to an integrated luminosity of $135~\ifb$.

The impact of hard-scatter primary vertex misidentification due to low track activity is strongly mitigated using a dedicated BDT discriminant. Novel data driven techniques are employed to estimate the background arising from electrons and jets misidentified as photons, while the normalisation of the background processes with a genuine photon is obtained using MC simulations constrained by data in dedicated control regions.

No significant excess over the SM expectation is observed. The results are interpreted in terms of a $H\to\gamma\gammad$ decay, yielding an observed (expected) 95\% CL upper limit on the branching ratio of $1.4\%$ ($1.2\%$) at 95\% CL, comparable to the best limit from previous searches, set by the ATLAS Run~2 combination. A combination with previous ATLAS Run 2 searches improves the observed (expected) limit to 0.9\% (0.9\%), representing the most stringent constraint to date on this decay mode.

% End of text imported from the .//sections/conclusion.tex input file

%
%
%
%
%
\section*{Acknowledgements}
%

%

% The next lines are included from the .//acknowledgements/Acknowledgements.tex input file
%
%

%
%

We thank CERN for the very successful operation of the LHC and its injectors, as well as the support staff at
CERN and at our institutions worldwide without whom ATLAS could not be operated efficiently.

The crucial computing support from all WLCG partners is acknowledged gratefully, in particular from CERN, the ATLAS Tier-1 facilities at TRIUMF/SFU (Canada), NDGF (Denmark, Norway, Sweden), CC-IN2P3 (France), KIT/GridKA (Germany), INFN-CNAF (Italy), NL-T1 (Netherlands), PIC (Spain), RAL (UK) and BNL (USA), the Tier-2 facilities worldwide and large non-WLCG resource providers. Major contributors of computing resources are listed in Ref.~\cite{ATL-SOFT-PUB-2026-001}.

We gratefully acknowledge the support of ANPCyT, Argentina; YerPhI, Armenia; ARC, Australia; BMWFW and FWF, Austria; ANAS, Azerbaijan; CNPq and FAPESP, Brazil; NSERC, NRC and CFI, Canada; CERN; ANID, Chile; CAS, MOST, NSFC and FRFCU, China; Minciencias, Colombia; MEYS CR, Czech Republic; DNRF and DNSRC, Denmark; IN2P3-CNRS and CEA-DRF/IRFU, France; SRNSFG, Georgia; BMFTR, HGF and MPG, Germany; GSRI, Greece; RGC and Hong Kong SAR, China; ICHEP and Academy of Sciences and Humanities, Israel; INFN, Italy; MEXT and JSPS, Japan; CNRST, Morocco; NWO, Netherlands; RCN, Norway; MNiSW, Poland; FCT, Portugal; MNE/IFA, Romania; MSTDI, Serbia; MSSR, Slovakia; ARIS and MVZI, Slovenia; DSI/NRF, South Africa; MICIU/AEI, Spain; SRC and Wallenberg Foundation, Sweden; SERI, SNSF and Cantons of Bern and Geneva, Switzerland; NSTC, Taipei; TENMAK, T\"urkiye; STFC/UKRI, United Kingdom; DOE and NSF, United States of America.

Individual groups and members have received support from BCKDF, CANARIE, CRC and DRAC, Canada; CERN-CZ, FORTE and PRIMUS, Czech Republic; COST, ERC, ERDF, Horizon 2020 and Marie Sk{\l}odowska-Curie Actions, European Union; Investissements d'Avenir Labex, Investissements d'Avenir Idex and ANR, France; DFG and AvH Foundation, Germany; Herakleitos, Thales and Aristeia programmes co-financed by EU-ESF and the Greek NSRF, Greece; BSF-NSF and MINERVA, Israel; NCN and NAWA, Poland; La Caixa Banking Foundation, CERCA and AGAUR programs from Generalitat de Catalunya and PROMETEO and GenT Programmes Generalitat Valenciana, Spain; G\"{o}ran Gustafssons Stiftelse, Sweden; The Royal Society and Leverhulme Trust, United Kingdom; Eric and Wendy Schmidt Fund for Strategic Innovation, United States of America.

In addition, individual members wish to acknowledge support from Chile: Agencia Nacional de Investigaci\'on y Desarrollo (ANID FONDECYT reg. 1230987, FONDECYT 1230812, FONDECYT 1240864, Fondecyt 3240661, Fondecyt Regular 1240721); China: Fundamental Research Funds for the Central Universities (010-63263105), Chinese Ministry of Science and Technology (MOST-2023YFA1605700, MOST-2023YFA1609300), National Natural Science Foundation of China (NSFC 12275265, NSFC-W2543005); Czech Republic: Czech Science Foundation (GACR - 24-11373S), Ministry of Education Youth and Sports (ERC-CZ-LL2327, FORTE CZ.02.01.01/00/22\_008/0004632); EU: H2020 European Research Council (ERC - 101002463); European Union: European Research Council (BARD No. 101116429, ERC - 101219398, ERC 101089007), European Regional Development Fund (HE COFUND GA No.101081355, ERDF), Marie Sklodowska-Curie Actions (GAP-101168829); France: Agence Nationale de la Recherche (ANR-21-CE31-0013, ANR-22-EDIR-0002, ANR-24-CE31-0504-01); Germany: Deutsche Forschungsgemeinschaft (DFG - 469666862); China: Research Grants Council (GRF); Italy: Istituto Nazionale di Fisica Nucleare (LHC-MIUR - 28003/2025), Ministero dell'Università e della Ricerca (NextGenEU I53D23000820006 M4C2.1.1, SOE2024\_0000023); Japan: Japan Society for the Promotion of Science (JSPS KAKENHI  JP25H0063, JSPS KAKENHI JP22H04944, JSPS KAKENHI JP24K23939, JSPS KAKENHI JP24KK0251, JSPS KAKENHI JP25H00650, JSPS KAKENHI JP25H01291, JSPS KAKENHI JP25K01011, JSPS KAKENHI JP25K01023, JSPS KAKENHI JP25KK0047); Poland: Polish National Science Centre (NCN OPUS 2023/51/B/ST2/02507, NCN OPUS nr 2022/47/B/ST2/03059, UMO-2019/34/E/ST2/00393, UMO-2021/42/E/ST2/00350, UMO-2022/47/O/ST2/00148, UMO-2023/49/B/ST2/04085, UMO-2023/51/B/ST2/00920, UMO-2024/53/N/ST2/00869); Spain: Agència de Gestió d'Ajuts Universitaris i de Recerca. (AGAUR - 2023 BP 00141), Ministry of Science and Innovation (RYC2019-028510-I, RYC2020-030254-I, RYC2021-031273-I, RYC2022-038164-I), Ministerio de Ciencia, Innovación y Universidades/Agencia Estatal de Investigaci\'on (EU NextGenerationEU (PRTR-C17.I1), PID2022-142604OB-C22); Sweden: Carl Trygger Foundation (Carl Trygger Foundation CTS 22:2312), Swedish Research Council (Swedish Research Council 2023-04654, VR 2021-03651, VR 2022-03845, VR 2022-04683, VR 2023-03403, VR 2024-05451, VR 2025-05940), Knut and Alice Wallenberg Foundation (KAW 2023.0366); United Kingdom: The Binks Trust, Royal Society (NIF-R1-231091); United States of America: U.S. Department of Energy (ECA DE-AC02-76SF00515), John Templeton Foundation (John Templeton Foundation 63206), Neubauer Family Foundation.

%
%

% End of text imported from the .//acknowledgements/Acknowledgements.tex input file

%
\clearpage
\printbibliography
\clearpage
% ATLAS Collaboration author list
% Reference date of HMBS-2024-50 is 2026-04-02
% Author list last updated on date 25-SEP-26
% Data extracted on 25-Sep-2026 for paper reference HMBS-2024-50
% at 10:00am
 
\begin{flushleft}
\hypersetup{urlcolor=black}
{\Large The ATLAS Collaboration}

\bigskip

\AtlasOrcid[0000-0002-6665-4934]{G.~Aad}$^\textrm{\scriptsize 102}$,
\AtlasOrcid[0000-0001-7616-1554]{E.~Aakvaag}$^\textrm{\scriptsize 17}$,
\AtlasOrcid[0000-0002-5888-2734]{B.~Abbott}$^\textrm{\scriptsize 121}$,
\AtlasOrcid[0000-0002-0287-5869]{S.~Abdelhameed}$^\textrm{\scriptsize 83b}$,
\AtlasOrcid[0000-0002-1002-1652]{K.~Abeling}$^\textrm{\scriptsize 54}$,
\AtlasOrcid[0000-0001-5763-2760]{N.J.~Abicht}$^\textrm{\scriptsize 48}$,
\AtlasOrcid[0000-0002-8496-9294]{S.H.~Abidi}$^\textrm{\scriptsize 30}$,
\AtlasOrcid[0009-0003-6578-220X]{M.~Aboelela}$^\textrm{\scriptsize 44}$,
\AtlasOrcid[0000-0002-9987-2292]{A.~Aboulhorma}$^\textrm{\scriptsize 36e}$,
\AtlasOrcid[0000-0001-5329-6640]{H.~Abramowicz}$^\textrm{\scriptsize 154}$
% ...
\bigskip
\\

% ...
$^{17}$Department for Physics and Technology, University of Bergen, Bergen; Norway.\\
% ...
$^{30}$Physics Department, Brookhaven National Laboratory, Upton NY; United States of America.\\
% ...
$^{36}$$^{(a)}$Facult\'e des Sciences Ain Chock, Universit\'e Hassan II de Casablanca;$^{(b)}$Facult\'{e} des Sciences, Universit\'{e} Ibn-Tofail, K\'{e}nitra;$^{(c)}$Facult\'e des Sciences Semlalia, Universit\'e Cadi Ayyad, LPHEA-Marrakech;$^{(d)}$LPMR, Facult\'e des Sciences, Universit\'e Mohamed Premier, Oujda;$^{(e)}$Facult\'e des sciences, Universit\'e Mohammed V, Rabat;$^{(f)}$Institute of Applied Physics, Mohammed VI Polytechnic University, Ben Guerir; Morocco.\\
% ...
$^{44}$Physics Department, Southern Methodist University, Dallas TX; United States of America.\\
% ...
$^{48}$Fakult\"{a}t Physik, Technische Universit{\"a}t Dortmund, Dortmund; Germany.\\
% ...
$^{54}$II. Physikalisches Institut, Georg-August-Universit\"{a}t G\"ottingen, G\"ottingen; Germany.\\
% ...
$^{83}$$^{(a)}$Khalifa University of Science and Technology, Abu Dhabi;$^{(b)}$New York University Abu Dhabi, Abu Dhabi;$^{(c)}$United Arab Emirates University, Al Ain;$^{(d)}$University of Sharjah, Sharjah; United Arab Emirates.\\
% ...
$^{102}$CPPM, Aix-Marseille Universit\'e, CNRS/IN2P3, Marseille; France.\\
% ...
$^{121}$Homer L. Dodge Department of Physics and Astronomy, University of Oklahoma, Norman OK; United States of America.\\
% ...
$^{154}$Raymond and Beverly Sackler School of Physics and Astronomy, Tel Aviv University, Tel Aviv; Israel.\\
% ...
$^{a}$ Also at Affiliated with an institute formerly covered by a cooperation agreement with CERN.\\
$^{b}$ Also at An-Najah National University, Nablus; Palestine.\\
$^{c}$ Also at Borough of Manhattan Community College, City University of New York, New York NY; United States of America.\\
$^{d}$ Also at Center for Interdisciplinary Research and Innovation (CIRI-AUTH), Thessaloniki; Greece.\\
$^{e}$ Also at Centre of Physics of the Universities of Minho and Porto (CF-UM-UP); Portugal.\\
$^{f}$ Also at CERN, Geneva; Switzerland.\\
% ...

\end{flushleft}

% Created with Glance <Atlas.Glance@cern.ch>

%

%
%

%

%
%
%
%

\end{document}